\documentclass[fleqn,usenatbib]{mnras}
\usepackage[T1]{fontenc}
\usepackage{graphicx}
\usepackage{amsmath}
\usepackage{amssymb}
\usepackage{newtxtext,newtxmath}
\usepackage{subfig}
\usepackage{float}
\usepackage{siunitx}
\DeclareSIUnit{\parsec}{pc}
\DeclareSIUnit{\arcsecond}{as}
\DeclareSIUnit{\year}{yr}
\DeclareMathOperator{\sech}{sech}
\newcommand{\scaleradius}{a}
\usepackage[normalem]{ulem}

\title[6-d Stellar Shells]{A 6-d view of stellar shells}

\author[C.~A. Dong-P\'aez et al.]{
C.~A. Dong-P\'aez,$^{1}$\thanks{Emails: chiandongpaez@gmail.com, eugvas@lpi.ru, nwe22@cam.ac.uk}
E. Vasiliev,$^{1,2}$
N.~W. Evans$^{1}$
\\
$^{1}$Institute of Astronomy, Madingley Rd, Cambridge CB3 0HA, UK\\
$^{2}$Lebedev Physical Institute, Leninsky Prospekt 53, Moscow, 119991, Russia}

\date{Accepted 2021 November 16. Received 2021 November 15; in original form 2021 October 3}

\begin{document}
\label{firstpage}
\pagerange{230--245}\volume{510}\pubyear{2022}
\setcounter{page}{230}
\maketitle

\begin{abstract}
Stellar shells are low surface brightness features, created during nearly head-on galaxy mergers from the debris of the tidally disrupted satellite. Here, we investigate the formation and evolution mechanism of shells in six dimensions (3d positions and velocities). We propose a new description in action-angle coordinates which condenses the seemingly complex behaviour of an expanding shell system into a simple picture, and stresses the crucial role of the existence of different stripping episodes in the properties of shells. Based on our findings, we construct a method for constraining the potential of the host galaxy and the average epoch of stripping. The method is applicable even if the shells cannot be identified or isolated from the data, or if the data are heavily contaminated with additional foreground stars. These results open up a new possibility to study the ancient merger that built the Milky Way Galaxy's stellar halo.
\end{abstract}



\section{Introduction}

In the picture of hierarchical galaxy formation, galaxies are built through constant accretion and mergers, which leave imprints on their morphology and dynamics. The tidal features resulting from the disruption of accreted satellite galaxies can take a wide range of forms depending on the properties of the progenitor mergers, including tidal streams, shells or umbrellas~\citep{He15}. In this work, we focus on the formation and properties of stellar shells.

Shells are roughly concentric low surface brightness arcs composed of stripped stellar material, whose existence was first reported by \citet{Arp66}. Observationally speaking, shell galaxies are classified into three classes according to the morphology of their shell systems \citep{Wilkinson87, Prieur90}. Type I shells lie along a single axis, and are interleaved in radius, with adjacent shells appearing on opposite sides of the galaxy; Type II shells are distributed randomly around the host; Type III shells are those irregular systems which do not resemble the two previous classes, usually because the shells are not concentric. The frequencies of the three categories are believed to be comparable \citep{Prieur90}. 

Stellar shells are common in the Local Universe, where such low surface brightness substructure can be resolved. Early studies noted that the incidence of shells in elliptical and lenticular galaxies locally is $\sim10\%$ \citep{Malin83, Schweizer85, Schweizer88, Atkinson13}. Shells have also been observed around spiral and dwarf galaxies \citep{Schweizer88, MartinezDelgado08, deBlok14}, although \citet{Atkinson13} claims that galaxies belonging to this population are half as likely to display shells as red galaxies. The incidence of shells has a significant environmental dependence. \citet{Malin83} found that shells are more common in galaxies that are isolated or in small groups. Recently, the incidence of shells around massive galaxies in the Illustris cosmological simulation \citep{Pop18} was estimated as $\approx 18\%$ at $z=0$.

Early optical studies by \citet{Sc80} on the giant elliptical NGC 1316 already suggested that shells may be caused by the infall of companions. The origin of shells as tidal debris from a past galaxy merger was first systematically investigated by \citet{Quinn84}. He examined N-body simulations of minor radial mergers of a low-mass disk with a spherical host potential. Subsequently, it became evident that the channels that can lead to shell formation are numerous. Several studies \citep[e.g.][]{Dupraz86, Hernquist88} revealed that shells can form under a wider set of conditions, including non-radial orbits, spheroidal satellites and high energy encounters. \citet{Hernquist92} showed that major mergers of two disc galaxies, in which both progenitors are completely destroyed and transformed, still can provide enough dynamically cold material to form shells. {\citet{Be85} provided an analytic framework of shells as caustics in cold collapse models}. More recently, the formation of shells has also been studied from a statistical perspective in the context of cosmological hydrodynamical simulations \citep{Pop18}. They find that shells that can be visually identifiable at $z=0$ are typically formed via radial mergers with a mass ratio $\gtrsim$ 1:10, in a lookback time window between $4-8 \, \si{\giga\year}$.

Early studies suggested that the distribution of shells can constrain the gravitational potential of the host galaxy and the timing of the original merger event \citep{Quinn84, Dupraz86, Canalizo07}. It has also been argued that the line-of-sight velocity distributions can be used to constrain the gravitational potential at the shell radii, the age of the shell system and the arrival direction of the progenitor \citep{Merrifield98, Ebrova12, Sanderson13}. Surface brightness models of shells, combining analytic models for the radial density profile and the phase-space distribution of shells, have also been explored as another means to constrain the potential \citep{Sanderson13}.

These results apply primarily to shells around external galaxies, for which individual stars can generally not be resolved and only projected positions and line-of-sight kinematics can be measured. Restricting the analysis to this three-dimensional space of projected positions and line-of-sight velocities severely reduces the information available. With the advent of the \textit{Gaia} satellite \citep{GAIA} and spectroscopic surveys such as \textsc{Weave} \citep{WEAVE}, \textsc{4most} \citep{4MOST} and \textsc{desi} \citep{DESI}, three-dimensional positions and velocities of Milky Way halo stars are becoming available out to distances of $100\,\si{\kilo\parsec}$. Very low surface brightness structures in the stellar halo are accessible to observation. The stellar halo is predominantly composed of stars deposited from disrupted satellite galaxies, and therefore it is a perfect laboratory to study the accretion history of our Galaxy, as well as its gravitational potential.

The goal of this investigation is to revisit the formation and evolution of shells from a numerical and analytical perspective, with stress on new insights that can be garnered from the six-dimensional phase-space positions of shell stars. In Section \ref{section:Simulations}, we introduce a suite of $N$-body simulations of shell systems formed in nearly radial minor mergers. We provide an overview of the properties of shells in phase space and action space, and provide new insights into the process of shell formation and evolution. Section~\ref{section:PropertiesOfShells} presents analytical formulae describing some fundamental properties of shells, focusing on the shell expansion velocity. In Section \ref{section:DeterminationOfPotential}, we propose and assess several new methods for constraining the potential and the timing of the event starting from simulated six-dimensional phase-space positions of shell particles. Finally, we summarise our results, examine the limitations of our method and outline possible further developments in Section \ref{section:Conclusions}.

\section{Simulations}
\label{section:Simulations}

\subsection{Numerical Details}
\label{subsection:NumericalMethod}

We model the host and satellite galaxies as spherically symmetric Navarro-Frenk-White (NFW) dark matter haloes \citep{NFW1997} with $10^5$ particles each. {This mean that the particle masses are smaller in the satellite, which aids resolution of the shell features.} The density profiles are exponentially truncated at the virial radius, obeying the following functional form,
\begin{equation}
    \rho (r) = \frac{\rho_0}{\frac{r}{\scaleradius}\left(1+\frac{r}{\scaleradius}\right)^2} \sech\left(\frac{r}{r_\text{vir}}\right),
\end{equation}
where $\scaleradius$ and $r_\text{vir}$ refer to the scale radius and the virial radius of the halo. Although pure gravitational simulations are scale-free, for the ease of comparison with the Milky Way system, we present our results in physical units: the mass of the host halo is fixed to ${M_\text{h} = 1.5 \times 10^{12} M_\odot}$, with virial radius ${r_\text{vir,h} = 200\, \si{ \kilo\parsec}}$ and scale radius ${r_\text{s,h} = 20\, \si{ \kilo\parsec}}$. 
Our $N$-body models represent dark haloes, but of course, shells are accumulations of stars. We use a particle-tagging method whereby a fraction of the dark matter particles of the satellite are chosen to represent the stars. Namely, stars are identified as the fraction $f_\text{mb} = 10\%$ \citep{LeBret17, He15} of the most bound particles in the satellite.

We vary two properties of encounter: the initial mass of the satellite halo $M_\text{s}$ and the initial circularity of its orbit $\eta$. This is defined as $\eta = J/J_\text{circ}(E)$, where $J$ is the angular momentum of the satellite and $J_\text{circ}(E)$ is the angular momentum of a circular orbit with energy $E$. 
We allow these quantities to vary in a $3\times3$ grid to give a total of nine simulations. Given that the main contributors to the Milky Way stellar halo are thought to have a mass ratio of $M_\text{s}/M_\text{h} \sim 1/10$ \citep{Bullock05}, we adopt the values $M_\text{s}/M_\text{h} = \{1/20, 1/10, 1/5\}$. The corresponding values of the scale radius of the satellite $r_\text{s,s}$ were selected according to the mass--concentration relation evaluated at redshift $z=0$ \citep[e.g.][]{Ludlow14}, namely $\scaleradius_\text{s}/\scaleradius = \{1/4, 2/5, 1/2\}$ for the corresponding mass ratios. Since the morphology of shells in configuration space depends strongly on the eccentricity of the encounter \citep{Hernquist88}, we let the initial circularity adopt the values $\eta = \{0, 0.2, 0.4\}$. For simplicity, we denote the three values of $\eta$ by \{R, I, M\}, which stand for Radial, Intermediate and Maximum, and the three values of $M_\text{s}/M_\text{h}$ by the subscripts $\{20, 10, 5\}$. This gives us a compact mnemonic to refer to any simulation -- for example, simulation $\text{R}_{20}$ is the one with $(M_\text{s}/M_\text{h}, \eta) = (1/20, 0)$. 
All simulations start with the satellite placed at a relative position $\mathbf{r_i} = (r_\text{vir,h}, 0, 0)$ from the host, with an initial velocity vector given by the circularity $\eta$ and contained in the ($x,y$) plane. Each simulation was run for $12 \,\si{\giga\year}$. 

The simulations were performed using the publicly available fast-multipole $N$-body code \textsc{GyrfalcON} \citep{GyrfalcON}. This code, along with other programs used in the creation and analysis of the simulations, are available in the stellar dynamics toolkit \textsc{Nemo} \citep{NEMO}. {We use the softening length $\epsilon = 0.5\, \si{ \kilo\parsec}$, but verified that the morphology of the debris and the disruption time does not depend on it.} The timestep was chosen as $\delta t = 0.1\epsilon/v_\text{rms}$, where $v_\text{rms}$ is the initial root-mean-square velocity of the simulated particles. This corresponds to $0.24 \, \si{\mega\year}$ and ensures that the fractional errors in the total energy and angular momentum are smaller than $1\times10^{-4}$.

In all nine simulations, the tidal disruption of the satellite leads to the production of shells.  Visually, shells generated by less massive progenitors extend to larger radii and appear to have sharper edges. In a spherical potential, radial encounters result in Type I shell systems, while shells produced by eccentric encounters are Type II as they lie along the different directions of the trajectory of the bound satellite, and can form on a significantly larger timescale due to the satellite surviving a larger number of pericentric passages.

It is sometimes convenient to compare different simulations at snapshots where shells are roughly uniformly developed. Due to the variety of orbital time scales and stripping times involved in a single simulation, this is not a straightforward task. We quantify the development of a set of shell systems by introducing a dimensionless time parameter $\tau$, given by the following simple recipe,
\begin{equation}
    \tau = \frac{t-t_\text{mode}}{T_{\text{mode}}}
\end{equation}
where $t_\text{mode}$ is the time of the pericentre passage that strips the highest number of particles and $T_{\text{mode}}$ is the orbital period of the satellite near $t_\text{mode}$. 

We will use the data extracted from the simulations in the rest of the paper. Before that, it is worth mentioning a few limitations. First, our simulations ignore the baryonic contribution to the host potential. However, the Milky Way is thought to be dark matter dominated at radii $r \gtrsim 14 \, \si{\kilo\parsec}$ \citep{Eilers19}. So, baryonic effects are only relevant for the very innermost shells, which our analyses generally ignore since their features are erased rapidly due to phase mixing. Although dark matter halo shapes are often complicated \citep[e.g.,][]{Bo16,Zavala19}, we use a spherical approximation to avoid introducing additional parameters. 
There are also limitations arising from the use of the particle-tagging technique, which relies on the assumption that stars follow the dynamics of the most bound dark matter particles. This assumption is clearly broken either if the gravity due to the stellar component is important (e.g. central stellar bulges) or if the morphology and kinematics of the stellar component differs from its halo (e.g. rotationally-supported stellar discs, \citealt{Amorisco17}). However, low-mass satellites considered in this work are dominated by dark matter at all radii and show little rotational support \citep{Walker09, Am11, Mayer01}, and therefore the use of the particle tagging method seems justified. We have also ignored the effects of varying the parameter $f_\text{mb}$, which regulates the size of the stellar component relative to the dark halo. In fact, our chosen value of $f_\text{mb}$ lies at the upper end of the acceptable range of values \citep{LeBret17}. However, due to the tight correlation of the initial energy with the stripping time, reducing $f_\text{mb}$ simply amounts to considering more initially tightly bound stars, and thus observing particles that are stripped later in time. 

\subsection{Shells in Phase Space}
\label{subsection:ShellsInPhaseSpace}

Shells are formed from stripped satellite stars, which are initially confined to a small region of phase space. Since our primary interest lies in the radial structure of the shell system and not in its orientation, we consider only the radial distance from the host centre $r$ and the corresponding velocity $v_r$. As stars orbit the host potential $\Phi(r)$ and phase wrap, the distribution in phase space is stretched, while maintaining a highly fine-grained phase-space density. Therefore, shells appear as nearly one-dimensional structures in the $(r,v_r)$ space, with small intrinsic scatter. For nearly radial encounters, stars stripped from the satellite subsequently move in the host-dominated potential on orbits similar to the progenitor orbit at the stripping time.

\begin{figure*}
    \centering
    \includegraphics[width=\textwidth]{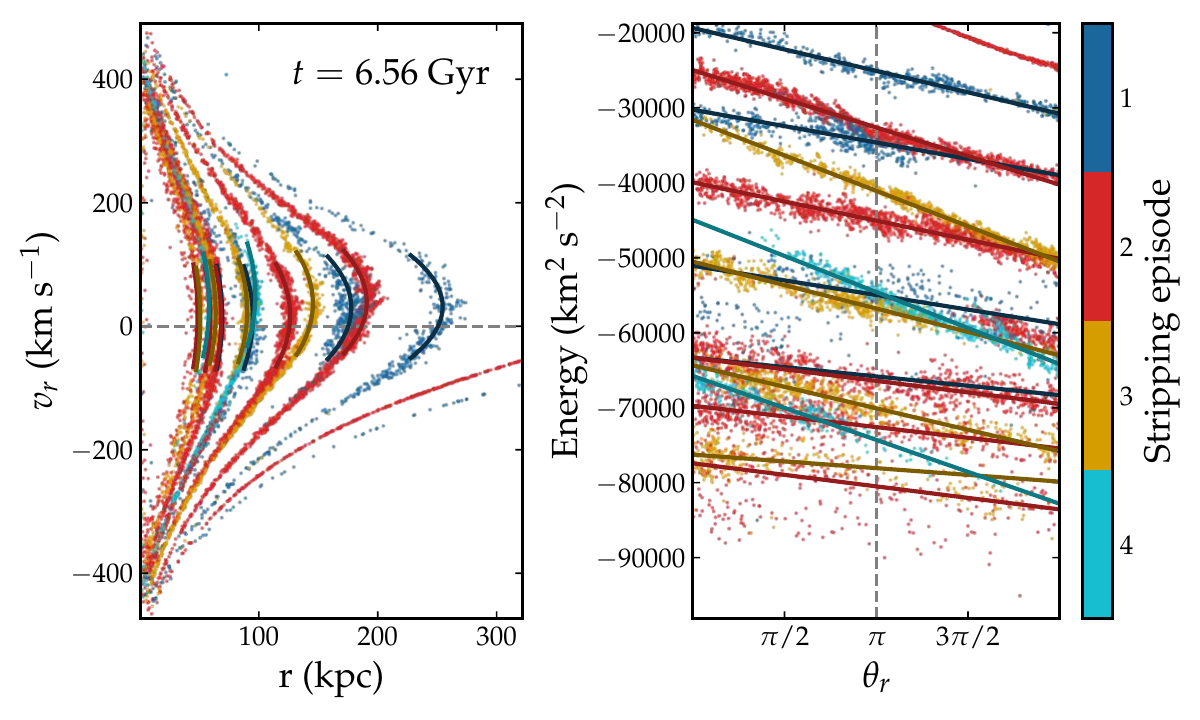}
    \caption{The distribution of shells in simulation $\text{R}_{20}$ at time $t = 6.56 \, \si{\giga\year}$ in phase space $(v_r, r)$ (left panel), and in energy-angle space $(E, \theta_r)$ (right panel). Lines in the left panel correspond to parabolas fitted to the particles located at the edge of each shell. In $(E, \theta_r)$ space, the shells are straightened out to follow nearly linear features. Lines in the right panel correspond to straight lines fitted to all particles in each shell. Particles are coloured according to their stripping time, where stripping episode $n$ corresponds to the time interval between the $n$th and $(n+1)$th satellite pericentre passages.}
    \label{fig:shells_phase+action_angle}
\end{figure*}

An example of the distribution of stripped stars in $(v_r, r)$ space is shown in the left panel of Fig.~\ref{fig:shells_phase+action_angle}. In this space, shells appear as bell-shaped curves. The outermost region of a shell can be approximated by a parabolic fit,
\begin{equation}
    r = r_\text{s} - \kappa(v_r - v_\text{s})^2.
    \label{eq:parabola}
\end{equation}
Here, $r_\text{s}$ is the maximum radius of the shell or shell radius, while $v_\text{s}$ is the radial velocity at $r_\text{s}$ (``shell velocity''). The fact that $v_\text{s}$ is positive shows the shell system is slowly expanding. Consequently, $v_\text{s}$ and $r_\text{s}$ can be measured by extracting the best-fit parameters of a parabolic fit to the outermost region of the shell. The parameter $\kappa$ is related to the gradient of the potential \citep{Sanderson13}
\begin{equation}
    \kappa \approx \frac{1}{2\Phi'(r_\text{s})},
\end{equation}
where the prime mark denotes a derivative with respect to $r$.

Despite the seeming simplicity of constructing a $(v_r, r)$ diagram, the properties of shells are far from simple. First, at fixed time, their distribution is not straightforward to model. Further, as noted already in early studies \citep[e.g.][]{Dupraz87}, the core of the progenitor can survive multiple pericentres, depositing stars at each passage, meaning that the stripping time of shell stars is not uniform in general. In this phase-space picture, observing the effect of shells having different stripping times is a difficult task. In absence of this information, the evolution of shells may seem complex: some properties, such as $v_\text{s}$, do not follow a one-to-one relation with the shell radius $r_\text{s}$. As shells can possess different velocities at the same radius, shells appear to cross, merge and divide.

\subsection{Shells in Action Space}
\label{subsection:ShellsInActionSpace}

The behaviour of shells can be more easily understood by performing a coordinate change $(\mathbf{x}, \mathbf{v}) \rightarrow (\boldsymbol{\theta}, \mathbf{J})$, from phase-space to action-angle coordinates. In this section, we introduce action-angle coordinates and describe the properties and evolution of shells in this space.

\subsubsection{Action-angle coordinates}

{Action-angle coordinates are discussed in many books on dynamics~\citep[e.g.,][]{LL,Go02,BinneyAndTremaine}.} They are a set of canonical coordinates in which the three momenta are integrals of motion. An integral of motion is a function of the phase space coordinates only (and not a function of time) which is constant along the orbit. Actions are particularly valuable as they are invariant under slow, adiabatic changes in the potential. Therefore, our findings regarding the properties of shells in action space will hold good, even if the potential is not time-independent, but changes through subsequent accretion and evolution. 

The actions in spherical polar coordinates are $\mathbf{J} = (J_r, J_\theta, J_\phi)$. The radial action is defined as
\begin{equation}
    J_r = \frac{1}{\pi}\int^{r_\text{apo}}_{r_\text{peri}} dr \, \sqrt{2E - 2\Phi(r) - \frac{L^2}{r^2}}
    \label{eq:radial_action}
\end{equation}
for a spherical, time-independent potential, where the energy $E$ and the angular momentum $L$ are conserved quantities. Here, $r_\text{apo}$ and $r_\text{peri}$ are the apocentric and pericentric distances of the orbit (the roots of the expression under the radical in Eq.~(\ref{eq:radial_action})). The azimuthal action $J_\phi$ and the longitudinal action $J_\theta$ are related to the components of the angular momentum $\mathbf{L} =(L_x,L_y,L_z)$. Specifically, $J_\phi = L_z$ and $J_\theta = L - |L_z|$, where $L = |\mathbf{L}|$.

The three corresponding canonically conjugate coordinates, called angles, $\boldsymbol{\theta} = (\theta_r, \theta_\theta, \theta_\phi)$ are defined by Hamilton's equations of motion, changing linearly with time:
\begin{equation}
    \dot{\theta_i} = \frac{\partial H}{\partial J_i} \equiv \Omega_i(\mathbf{J}),
\end{equation}
where $\Omega_i$ are corresponding frequencies. As discussed above, we are mainly concerned with the radial motion, hence $J_r, \theta_r$ and $\Omega_r$ are the most relevant quantities for our problem.
The radial frequency $\Omega_r$ is inversely proportional to the radial orbital period $T_r$, and for high-eccentricity orbits ($L\approx 0$), can be approximated as a function of energy alone:
\begin{equation}
    \Omega_r(E) = \frac{2\pi}{T_r(E)},\qquad
    T_r(E) = 2\int^{r_\text{apo}(E)}_{0}\frac{dr}{\sqrt{2(E-\Phi(r))}}.
    \label{eq:radial_frequency}
\end{equation}
In any realistic potential, $\Omega_r$ is lower for high-energy (least bound) orbits. Furthermore, for nearly radial orbits, $J_r \gg L$, and we may label the particles by $J_r$, $E$ or $r_\text{apo}$ interchangeably.

By convention, a value of the radial angle $\theta_r = 0$ corresponds to pericentre, while $\theta_r = \pi$ corresponds to apocentre. 
Therefore, the time evolution of the radial angle of a given particle can be described as
\begin{equation}
    \theta_r = \Omega_r\, \Delta t,
    \label{eq:theta_evolution}
\end{equation}
where $\Delta t$ is the time since the particle was stripped at pericentre. 
A whole radial period is completed for every $2\pi$ interval, and therefore all $\theta_r$ separated by $2\pi$ are physically equivalent. To avoid ambiguity, we restrict $\theta_r$ to the interval $[0, 2\pi)$ where necessary.

The transformation from the positions and velocities to action-angle coordinates is performed with the publicly available stellar dynamics package \textsc{Agama} \citep{AGAMA}. Positions and velocities of particles are calculates with respect to the position and velocity of the centre of the bound host. Actions and angles are calculated using an NFW fit to the stellar potential, while particle energies are taken directly from the $N$-body snapshot.

\subsubsection{Shell Formation and Evolution in Action Space}
\label{subsubsection:ShellFormationAndEvolution}

The evolution of shells is beautifully simple in action space (Fig.~\ref{fig:shells_phase+action_angle}, right panel). Stars are stripped in a series of episodes, which occur near pericentre passages of the bound satellite. Particles that have been stripped from the satellite and move in the nearly time-independent host potential $\Phi(r)$ conserve their energy $E$. Therefore, stripped particles only move horizontally in the $(E, \theta_r)$ space, at constant speed in $\theta_r$ given by the radial frequency $\Omega_r(E)$. In reality, the potential does vary with time due to the accretion of the satellite and the deformation of the host galaxy by the satellite perturbations. However, this time-dependent effect on the conservation of energy is weak even for the highest satellite mass considered, and negligible after the satellite has been disrupted.

Consider a set of stars that are stripped near the $n$th pericentric passage of the progenitor at $\theta_r = 0$. As the progenitor approaches pericentre, the energy spread of the soon-to-be stripped particles is increased, stretching their distribution in the vertical direction. This effect can be understood in simple terms as follows. The distribution of bound satellite particles has a small velocity dispersion $\sigma_v$ arising from the gravity of the bound halo. In the frame of the host halo, the satellite also moves coherently at speed $\bar{v}$. The kinetic energy dispersion of bound particles is of the order ${\sim (\bar{v} + \sigma_v)^2/2 - \bar{v}^2/2 \approx \bar{v}\sigma_v}$. Hence, the energy dispersion is maximised at pericentre. Since most particles are stripped at or close to pericentre, this enlarged energy dispersion becomes imprinted in the distribution of the stripped particles. Denoting the time since the stripping episode by $\Delta t$, then, just after $\Delta t = 0$, all recently stripped particles lie on a vertical line at $\theta_r = 0$, with a wide initial energy distribution.

Since $\Omega_r$ increases monotonically with decreasing $E$ (towards more tightly bound orbits), particles with lower energy move faster horizontally in the ($E,\theta_r$) space. This horizontal shear causes the initially vertical line to tilt and its gradient decreases over time. At any given time, the gradient is shallower for more bound stars. Stars with lower energy also complete a larger number of orbits, wrapping up in $\theta_r$. 
The number of shells resulting from this single stripping episode can then be identified with the number of wraps of this line, which increases with time.

\begin{figure*}
  \centering
  \includegraphics[width=\textwidth]{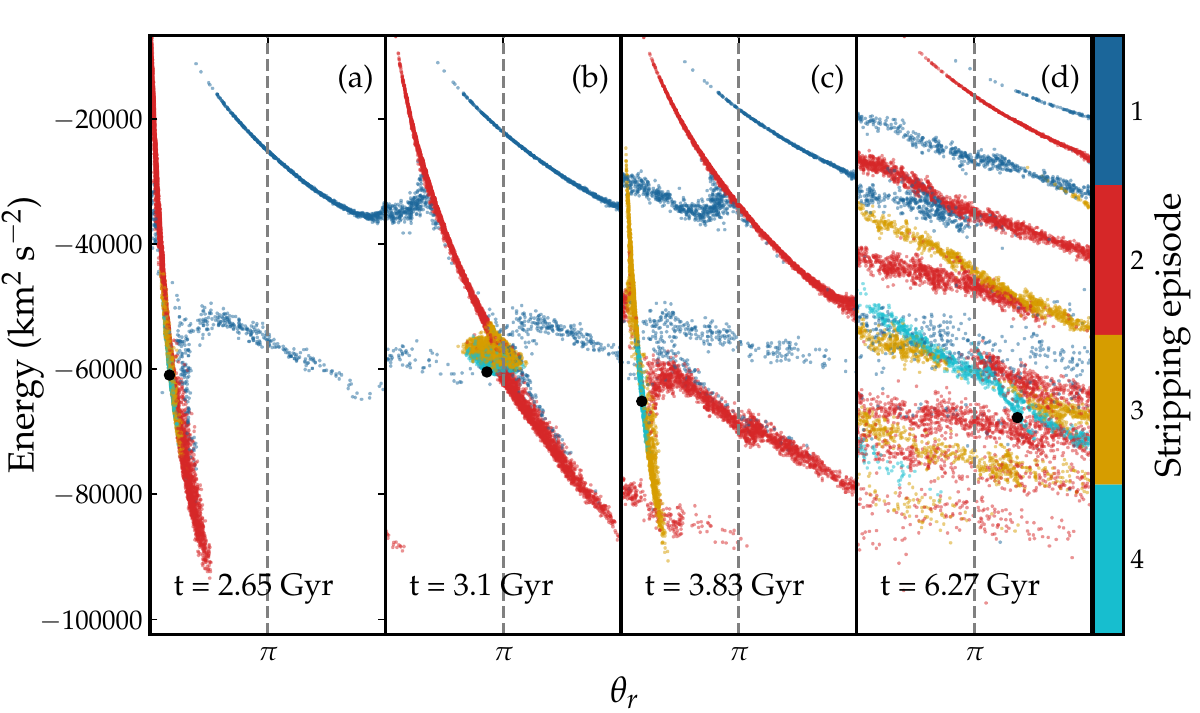}
  \caption{The evolution of the shell system corresponding to the second stripping episode in Simulation $\text{R}_{20}$. {The horizontal axis records the radial angle $\theta_r$, which is zero at pericentre and $\pi$ at apocentre.} Coloured points represent stars, with the colour indicating the episode at which each star has been stripped. The black dot represents the location of the bound satellite.  From left to right: (a) Particles belonging to the second stripping episode increase in energy dispersion and start as a vertical line at the stripping time, $\Delta t = 0$. (b) The particles evolve in angle, with $\Omega_r(E)$ increasing for more negative energies. This produces a horizontal shearing deformation which progressively stretches the line and decreases its gradient. (c) Material phase wraps, creating shells. An additional pericentre passage occurs, depositing further material and opening a gap in the original distribution of particles. Some 2nd-episode particles near the base of the tidal tails are perturbed by the gravity of the bound satellite (d) After sufficient horizontal stretching and phase wrapping, individual shells approach a linear relation with angle. Subsequent stripping episodes open further gaps.}
  \label{fig:shell_evolution}
\end{figure*}

As illustrated in Fig.~\ref{fig:shell_evolution}, this occurs whenever the bound satellite approaches pericentre. Therefore, every pericentric passage contributes a set of shells, which can be described by the time of its pericentre passage, the initial energy distribution and the spread in angle. The average energy of particles stripped in subsequent episodes decreases with time as dynamical friction causes the satellite to sink deeper into the host potential. As stars conserve their energy in a time-independent potential and therefore only evolve in $\theta_r$, the shells belonging to the same stripping episode do not (usually) overlap in energy.

However, due to the broadening near pericentre, the extremes of the initial energy distribution of a particular episode may sometimes overlap with the energy of the preceding episodes, and so the resulting shells belonging to different episodes can also overlap in $E$ and $r_\text{s}$. Therefore, some properties of shells may not be one-to-one with respect to radius, as they may depend on the stripping time. This explains the shell crossing events observed in $(v_r, r)$.

\begin{figure*}
  \centering
  \includegraphics[width=\textwidth]{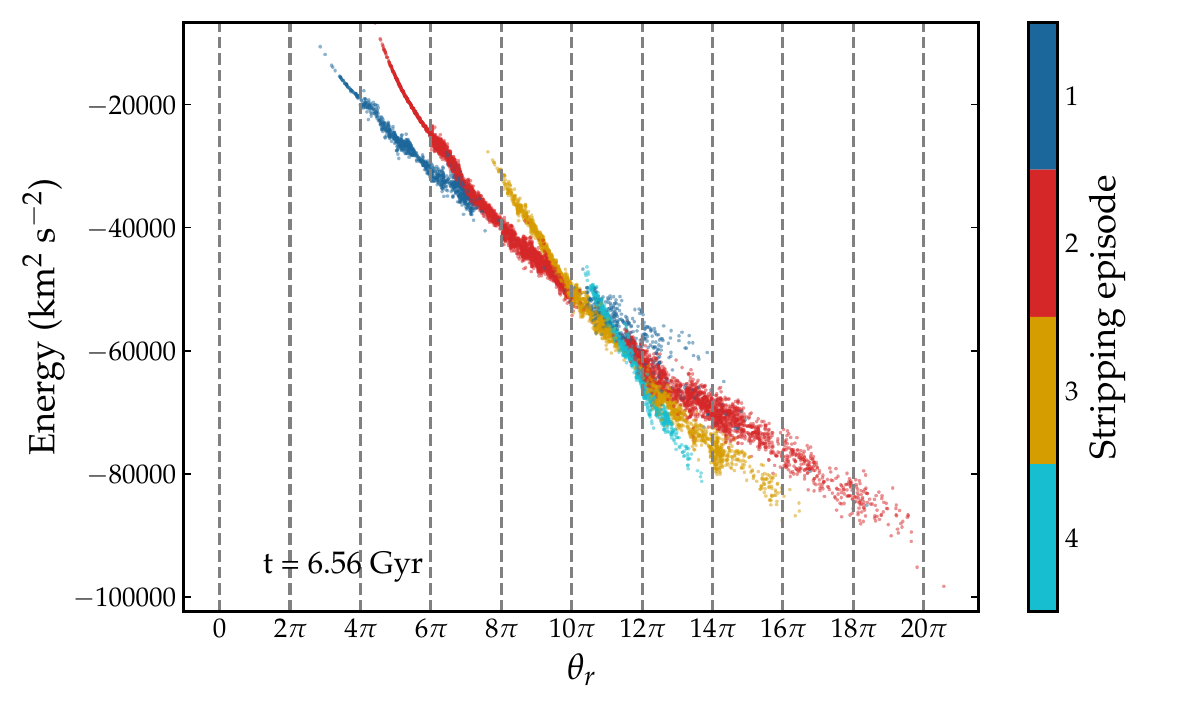}
  \caption{The distribution of star particles in $(E,\theta_r)$ space, for the same snapshot shown in Fig.~\ref{fig:shells_phase+action_angle} (simulation $\mathrm{R}_{20}$, $t = 6.56 \, \si{\giga\year}$), where the angles have been unfolded to their full value (i.e. not modulo $2\pi$). Pericentre passages, occurring at integer multiples of $2\pi$, are indicated with dashed vertical lines. Colours encode the episode at which each particle is stripped. Particles belonging to the $n$th episode are stripped at the $n$th pericentre (which takes place at $\theta_r=2n\pi$). In this space, all particles belonging to the same stripping episode lie on a single curve.}
  \label{fig:shells_unfolded}
\end{figure*}

Fig.~\ref{fig:shells_unfolded} provides perhaps a more illustrative picture of the evolution of shells in $(E,\theta_r)$ space. This figure is equivalent to the right panel of Fig.~\ref{fig:shells_phase+action_angle}, but here the full unwrapped value of $\theta_r$ (i.e. not modulo $2\pi$) is shown on the horizontal axis. In this space, all particles belonging to the same episode lie on a single (but not necessarily continuous) curve with a decreasing slope. As mentioned earlier, particles belonging to the $n$th episode are stripped at the $n$th pericentre passage, which occurs at $\theta_r = 2n\pi$. Each such set of particles starts as a vertical line at the stripping episode $\theta_r = 2n\pi$. This line is subsequently sheared horizontally due to the energy-dependence of the orbital frequency $\Omega_r$ into a curve with varying gradient. The curves corresponding to different stripping episodes partially overlap in energy, and have different gradients, since earlier episodes have undergone a larger amount of shearing. In practice, $\theta_r$ can only be determined observationally up to modulo $2\pi$, so each curved line will be usually wrapped and thus observed as set of lines (shells) with slowly varying gradient spanning an angle of $2\pi$ (see Fig.~\ref{fig:shells_phase+action_angle}, right panel).

This simple picture is complicated by effects arising from approximating the potential as smooth, spherically symmetric and time-independent, and neglecting the reflex motion and deformation of the host galaxy in response to the satellite. First, the gravitational attraction by the bound satellite opens a gap in energy between the leading and trailing arms, implying that if the progenitor is not yet disrupted, the particles stripped at a given pericentre do not form a single continuous line in $(E, \theta_r)$. The location energy of the gaps is controlled by the energy of the satellite at the corresponding pericentre passages, while the width is controlled by the mass of the satellite at pericentre. Second, near pericentre, due to the higher concentration of host particles and the arrival of other shell particles, effects that alter the individual energies of particles -- such as 2-body encounters -- may be significant. This increases the intrinsic scatter in $E$ of particles, smearing the fine-grained structure of shells. Finally, considering particles stripped at the same episode, although most are  stripped immediately after their corresponding pericentre passage, a few, located near the base of the tidal tails, are either recaptured by the satellite when its tidal radius increases as it moves away from the pericentre, or remain marginally bound and are progressively stripped soon after, before the next pericentre passage. The latter population can be influenced by the gravity of the satellite during their first orbit, especially for massive progenitors. This effect can be observed in Fig.~\ref{fig:shell_evolution} -- shells are sometimes perturbed near the location of the bound satellite. Nonetheless, the interaction is generally weak: most of the particles behave as if they had been effectively stripped at the episode's pericentre passage and any small perturbations are diluted as phase wrapping proceeds.

The details of the process may also depend on the properties of the merger. For satellites with higher mass, the initial energy distributions of different stripping episodes do not overlap much with those of previous episodes.
A similar, but much weaker, trend is observed with increasing circularity $j$, due to the fact that the pericentre passages occur at larger distances, leading to weaker interactions. This enables the satellite to survive a greater number of pericentre passages, thus producing a higher number of shell systems. Dynamical friction has a stronger effect on satellites with higher $M_\text{s}$, for which the orbital energy is dissipated in a shorter timescale and the encounter proceeds faster. The increase in the overall potential due to accretion is also relevant: stars phase-mix on a shorter timescale for higher $M_\text{s}$.

\subsubsection{An Algorithm for the Identification and Isolation of Shells}
\label{subsubsection:IdentificationAndIsolation}

In the light of our qualitative discussion of shell formation and evolution, we motivate a simple but robust method in order to identify members of simulated shells.

Shell properties depend strongly on the stripping episode that originated it. Consequently, it is convenient to separate the stars in the simulation according to their stripping episode. In each particular snapshot, we tag stars as bound or unbound to the satellite according to the following iterative algorithm:

\begin{enumerate}
  \item Select all particles initially belonging to the satellite and compute the potential due to their self-gravity. Find the most bound particle and take the velocity of the halo $\mathbf{v_\text{sat}}$ to be the velocity of this particle. Assume initially that all satellite particles are bound.
  \item Calculate the velocity of every particle relative to $\mathbf{v_\text{rel}}$, as $\mathbf{v}_\text{rel} = \mathbf{v} - \mathbf{v_\text{sat}}$.
  \item Calculate the total specific energy $E_\text{sat}$ of every particle, as $E_\text{sat} = \frac{1}{2} |\mathbf{v}_\text{rel}|^2 + \Phi_\text{bound}$, and identify the new set of bound particles, satisfying $E_\text{sat}<0$.
  Recompute the potential only due to the gravity of the bound particles at the current iteration, $\Phi_\text{bound}$.
  \item Set $\mathbf{v_\text{sat}}$ to the average velocity of bound particles. Repeat the process from step (ii).
\end{enumerate}

After convergence, bound particles are identified as those with negative $E_\text{sat}$. Additionally, the velocity of the bound satellite halo is taken as $\mathbf{v_\text{sat}}$, while the bound mass of the halo is the total mass of the bound particles. Further, we identify the position of the satellite with the position of the most bound particle. The algorithm converges quickly, with at most four iterations typically required. We note that the host potential and the tidal radius of the satellite are not used in this analysis; nevertheless, particles outside the tidal radius usually have high relative velocities and hence are classified as unbound anyway.

Using this algorithm, each particle can be assigned a stripping time $t_\text{strip}$, the first time at which the particle is identified as unbound. Motivated by the previous discussion, we can also assign a time interval corresponding to each stripping episode. The $n$th stripping episode is defined as the time interval between the $n$th and $(n+1)$th satellite pericentric passages. Particles corresponding to the $n$th-episode behave as having been stripped almost exactly at the $n$th pericentre, starting initially as a vertical line in $(E, \theta_r)$ space. As special cases, the $0$th stripping episode is defined as the time before the first pericentre passage, while the last stripping episode is the time between the last pericentre passage and the moment of disruption. The satellite is considered to be completely disrupted when all particles have been identified as unbound in at least one snapshot. This method works well for all simulations, but it is optimal under the condition that the contribution to the self-gravity of the stripped particles is small -- that is, for low-mass, high-concentration halos.

\begin{figure*}
  \centering
  \includegraphics[width=\textwidth]{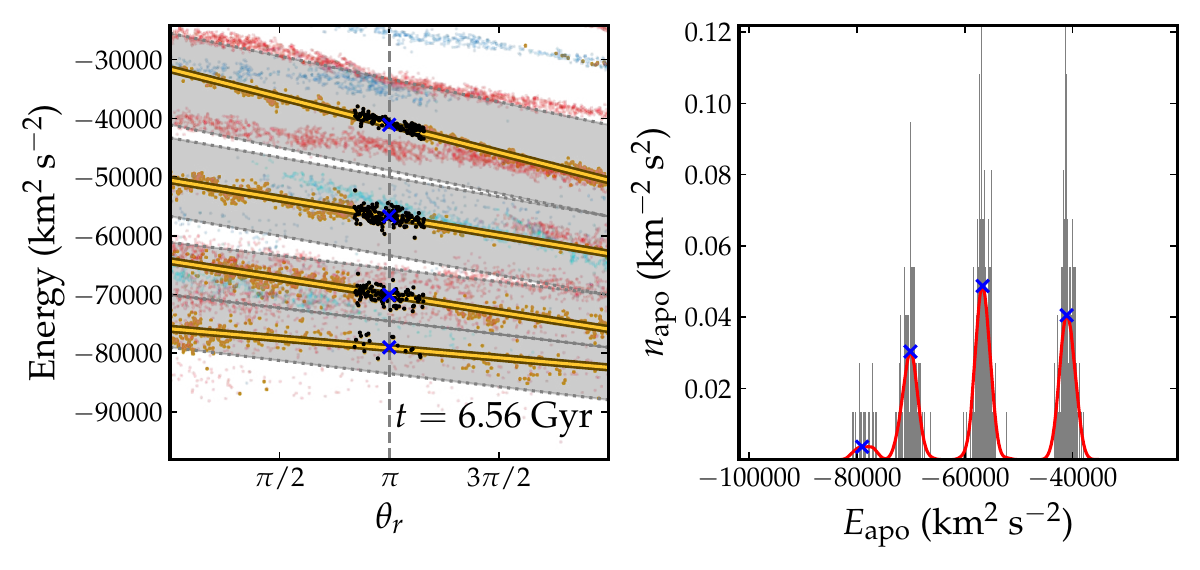}
  \caption{The algorithm for identification and isolation of shells, sketched for the $3$rd stripping episode in Fig.~\ref{fig:shells_phase+action_angle}. Left panel: The distribution of $3$rd-episode particles in $(E, \theta_r)$ space (orange dots). Particles highlighted in black satisfy $|\theta_r-\pi| < \pi/6$, and are selected for shell identification (right panel). Blue crosses denote the $E_\text{apo}$ peaks identified in the right panel. Grey regions enclose the particles isolated as constituents of each shell. Right panel: The distribution of the black particles highlighted in the left panel, in grey. The red line indicates the smooth distribution after applying a Gaussian filter, and the blue crosses mark the density peaks of this distribution, each corresponding to the location of an individual shell shown in the left panel.}
  \label{fig:shell_identification}
\end{figure*}

Next, we establish an algorithm for identifying shells and isolating their constituent particles, which is sketched in Fig.~\ref{fig:shell_identification}. We rely on the key observation that shells belonging to the same stripping episode do not overlap. It follows that, if only the particles belonging to a given episode in a narrow $\theta_r$ window are selected, the energy distribution of the particles is sharply peaked. Each peak corresponds to a particular shell crossing the $\theta_r$ window. Therefore, to find the energy $E_\text{apo}$ at which $n$th-episode shell lines cross $\theta_r = \pi$, we apply the following algorithm:
\begin{enumerate}
    \item Select the particles in the narrow interval around the apocentre $\theta_r \in (\pi-\delta\theta_r, \pi+\delta\theta_r)$. We use $\delta\theta_r = \pi/6$.
    \item Find the number density of these particles per unit energy, applying a kernel density estimation.
    \item Find the location of peaks in $n_\text{apo}$. The corresponding values of $E$ can be identified with the $E_\text{apo}$ of the shells. Similarly, we identify the local inter-shell $E_\text{apo}$ spacing, $\Delta E_\text{apo}$, as the minimum distance in $E_\text{apo}$ to a neighbouring shell.
\end{enumerate}

Once $E_\text{apo}$ of all $n$th-episode shells are known, their constituent particles can be isolated. Note that, if $\theta_r$ were not restricted to $(0, 2\pi)$ and ignoring the gap between leading and trailing arms, all $n$th-episode shells lie along a single line and the gradient of shell lines does not vary dramatically across neighbouring shells for a reasonably evolved system. Therefore, for a particular shell with apocentric energy $E_\text{apo}$, by imposing continuity with neighbouring shells on the shell boundaries and assuming constant spacing between shells, its $\Delta E_\text{apo}$ must be related to the average shell gradient. Shell lines approximately follow
\begin{equation}  \label{eq:shell_energy_gradient}
    E_\text{line}(\theta_r;\Delta E_\text{apo}) = -\frac{\Delta E_\text{apo}}{2\pi}(\theta_r-\pi) + E_\text{apo}
\end{equation}
Hence, for this shell, we can tag its constituent particles as those lying in the $(E,\theta_r)$ plane enclosed by the two lines $E_\text{line}(\theta_r;\Delta E_\text{apo}) - \tfrac{1}{2}\Delta E_\text{apo}$ and $E_\text{line}(\theta_r;\Delta E_\text{apo}) + \tfrac{1}{2}\Delta E_\text{apo}$. This interval is large enough such that no selection bias is introduced.

After all shells have been identified and their constituents have been isolated, refined estimates of $E_\text{apo}$ and $\partial E / \partial \theta_r$ can be obtained by fitting straight lines. Complementary to the action--angle space analysis, we also measure the radius and velocity of each shell in its apocentre $v_\text{s}$ and $r_\text{s}$ by fitting parabolas to the outermost regions of the shells. The result of this procedure is shown in the right panel of Fig.~\ref{fig:shells_phase+action_angle}. The multiple shells corresponding to each stripping episode have been reduced to a single continuous curve in ($E,\theta-r$) on identifying $0$ with $2\pi$. This algorithm is very robust. Its results are only unreliable when severe phase mixing has caused the inter-shell separation $\Delta E_\text{apo}$ to be smaller than the shell intrinsic scatter. This obstacle cannot be overcome except by incorporating detailed information about the timing of the event.

When dealing with observational data, the stripping episode to which each particle corresponds is, of course, not known {\it a priori}, and so this method does not apply.

\section{The Expansion of the Shell System}
\label{section:PropertiesOfShells}

In $(v_r, r)$ space, the shells have the property that the maximum radius $r_\text{s}$ occurs at positive values of $v_r$ rather than at $v_r = 0$. At outset, this is a little surprising. If shells were composed of a set of monoenergetic particles on almost radial orbits, it is clear that, as particles slow down and reverse direction at apocentre, $r_\text{s}$ would indeed correspond to $v_r = 0$. This is not the case.

The phenomenon of the expansion of the shell system is caused by the negative energy gradient of shells in $(E, \theta_r)$ space. As particles evolve in time toward increasing $\theta_r$, the particles at any given $\theta_r$ at some instant are replaced in the next instant by particles with higher energy according to the instantaneous energy gradient. Since all orbits have a similar morphology (all are roughly radial), this can be directly translated to an increase in radius, and therefore to a coherent expansion of the shell system. A shell can only be stationary if the overall energy gradient is zero, or equivalently, if the energy distribution of its particles is independent of $\theta_r$. 

To derive an approximate expression for the shell velocity $v_\text{s}$, we assume that shells lie along curves in $(E, \theta_r)$ space with negligible intrinsic scatter and that the stripped particles move in a smooth, time-independent host potential $\Phi(r)$. We label shell particles by the energy $E$, and make a continuum approximation, whereby shells are identified with a one-dimensional region of phase space (a curve in $(v_r, r)$ space peaking at $v_\text{s}$ with a value of $r = r_\text{s}$) or action space (a nearly straight line in $(E, \theta_r)$ space).
Therefore, along shell lines at fixed time, $\theta_r$, $v_r$ and $r$ can all be considered functions of $E$ only. 

The non-zero shell velocity can be approximately identified with the net motion of the shell outer boundary at $r_\text{s}$ \citep[e.g.][]{Quinn84, Ebrova12},
\begin{equation}
    v_\text{s} \approx \frac{dr_\text{s}}{dt}.
    \label{v_r0_expansion}
\end{equation}
This equation can be expanded to first order as
\begin{equation}
\label{eq:vs}
    v_\text{s} \approx -\left[\frac{\partial r}{\partial E}\right]_t \left[\frac{\partial E}{\partial \theta_r}\right]_t \left[\frac{\partial \theta_r}{\partial t}\right]_E.
\end{equation}
where every term is evaluated at $E=E_\text{s}$, the energy corresponding to the particles instantaneously at radius $r_\text{s}$. Note that the first and second terms are partial derivatives at fixed time \textit{along the shell line}, that is, along the one-dimensional region corresponding to the shell, which is parametrised by $E$, while the third term is a time-derivative at fixed $E$ (or equivalently, at fixed particle). The second and third terms together are $dE_\text{s}/dt$. Intuitively, this can be understood as follows. $E_\text{s}$ is the energy of the shell elements -- or particles -- which are instantaneously located at radial angle $\theta_s$. Now, $\theta_s$ is the angle corresponding to elements with $v_r = v_\text{s}$, which, provided that $v_\text{s}$ is sufficiently small, is close to $\pi$ and varies slowly. Shell elements in the vicinity of $E_\text{s}$ move locally with a coherent frequency $\left[\partial \theta_r / \partial t\right]_E$, and have a negative instantaneous local energy gradient $\left[\partial E / \partial \theta_r\right]_t$. Therefore, after a small time increment $\delta t$, shell elements are displaced to higher $\theta_r$, and so particles originally at radial angle $\theta_s$ are replaced by particles with higher energy due to the negative energy gradient. Consequently, $E_s$ is increased as the energy difference between the particles at $\theta_s$ before and after $\delta t$ is ${-\left[\partial E / \partial \theta_r\right]_t \left[\partial \theta_r / \partial t\right]_E \delta t}$.

Provided $v_\text{s}$ is sufficiently small, the particles in the vicinity of $E_\text{s}$ will be close to their apocentre, and therefore the change in their kinetic energies can be neglected compared to that in their potential energies. Hence,
\begin{equation}
    \left[\frac{\partial E}{\partial r}\right]_t \approx \frac{d\Phi}{dr}.
    \label{eq:energy_approximation}
\end{equation}
The term $[\partial E/\partial \theta_r]_t$ corresponds to the energy gradient of the shell at $E_\text{s}$. Henceforth, for notational simplicity, we drop the brackets and subscript $t$ from this term. The last term in eq.~(\ref{eq:vs}) corresponds to the frequency of particles with $E=E_\text{s}$, denoted by $\Omega_r(E_\text{s})$. Putting everything together, we obtain
\begin{equation}
    v_\text{s} = -\frac{\Omega_r(E_\text{s})}{\Phi'(r_\text{s})} \frac{\partial E}{\partial \theta_r}.
    \label{eq:v_r0_intermediate}
\end{equation}
Note that this does not assume radial orbits, and so eq.~(\ref{eq:v_r0_intermediate}) is valid for non-radial orbits provided their morphology is locally similar. The energy gradient $\partial E/\partial \theta_r$ at time $t$ can also be computed from the properties of the potential and the stripping episode.

Particles belonging to the same shell are stripped at lookback time $\Delta t$ near a satellite pericentric passage. Particles therefore start as a vertical line in $(E, \theta_r)$ space, which then shears horizontally. Recalling the time evolution of $\theta_r$ for a single particle, we can differentiate eq.~(\ref{eq:theta_evolution}) with respect to $E$ to obtain
\begin{equation}
    \frac{\partial E}{\partial \theta_r} = -\frac{T_r^2(E)}{2\pi \Delta t} \frac{dE}{dT_r},
    \label{eq:energy_gradient}
\end{equation}
where we have used eq.~(\ref{eq:radial_frequency}). By differentiating the radial period with respect to energy, we obtain
\begin{equation}
    \frac{dT_r}{dE} = \frac{T_r}{\Phi'(r_\text{apo}) r_\text{apo}} - \frac{r_\text{apo}}{\sqrt{2}}\int^{1}_{0}\frac{\left(1-x\frac{\Phi'(x\,r_\text{apo})}{\Phi'(r_\text{apo})}\right)dx}{\big[E-\Phi(x\,r_\text{apo})\big]^{3/2}}.
    \label{eq:dT_dE}
\end{equation}
Therefore, at time $\Delta t$, shells belonging to the same stripping episode tend to lie along lines with an energy gradient which depends on the potential and $\Delta t$, as described in eq.~(\ref{eq:energy_gradient}). In this form, eq.~(\ref{eq:dT_dE}) does assume radial orbits for $T_r$. However, the dependence of $T_r$ on the eccentricity of the orbit is relatively weak and the expressions could be generalised. Finally, introducing eq.~(\ref{eq:energy_gradient}) into eq.~(\ref{eq:v_r0_intermediate}), we obtain
\begin{equation}
    v_\text{s} = \frac{2\pi}{\Phi'(r_\text{s})\, \Omega_r(E_\text{s})\, \Delta t} \frac{dE}{dT_r}\bigg|_{E_\text{s}}.
    \label{eq:v_r0_final}
\end{equation}
Therefore, the shell velocity can be calculated from $E_\text{s}$, together with the potential $\Phi(r)$ and the stripping time $\Delta t$. 

To gain insight, we remark that the first term on the right hand side of eq.~(\ref{eq:dT_dE}) dominates for small $r_\text{s}$ (tightly bound shells). Neglecting the second term is equivalent to assuming that the variation in $T_r$ is proportional to the change in the path length of the orbit. Under this approximation, eq.~(\ref{eq:v_r0_final}) simplifies to
\begin{equation}
    v_\text{s} \approx \frac{r_\text{s}}{\Delta t}.
    \label{eq:v_r0_approximation}
\end{equation}
which gives an approximate scaling of $v_\text{s}$. This equation is independent of the potential, which implies that if eq.~(\ref{eq:v_r0_intermediate}) or (\ref{eq:v_r0_final}) are used to constrain the potential, data from shells at large radii generally contain more information than shells closer to the centre of the host galaxy. 

\begin{figure}
    \centering
    \subfloat[Simulation $\text{R}_{20}$, $t = 6.56 \, \si{\giga\year}$]{\includegraphics[width=0.5\textwidth]{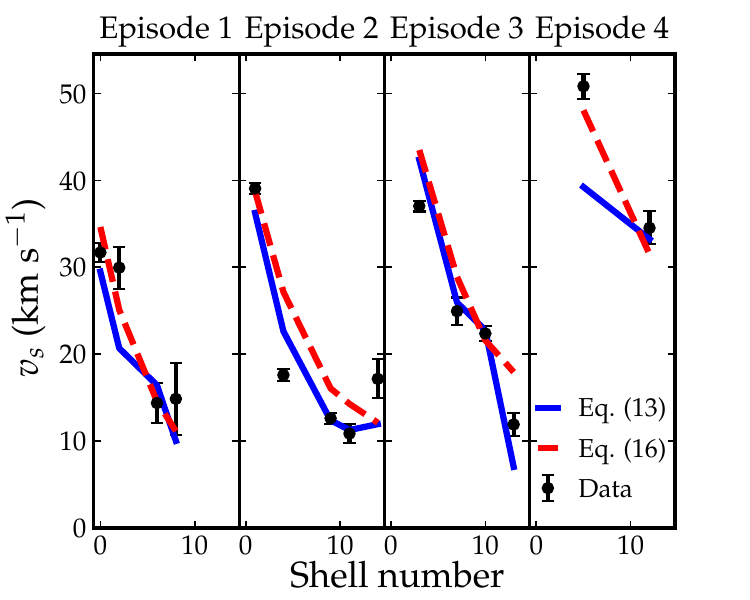}}
  \hfill
  \subfloat[Simulation $\text{M}_{20}$, $t = 10.98 \, \si{\giga\year}$]{\includegraphics[width=0.5\textwidth]{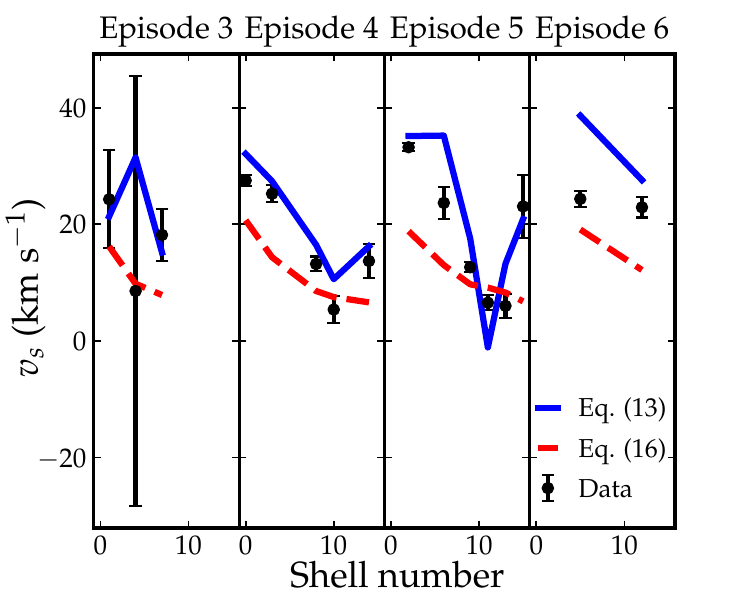}}
    \caption{The shell velocity against the shell number, ranked by decreasing total energy, for two different snapshots. The black errorbars display the measured values, obtained by fitting parabolas to the outer regions of shells. The blue solid line represent the value predicted by eq.~(\ref{eq:v_r0_intermediate}), which takes as input the measured energy gradient $\partial E / \partial \theta_r$ and the host potential, and the red dashed line the prediction from eq.~(\ref{eq:v_r0_final}), which takes as input the stripping time of the shell and the host potential.}
    \label{fig:v_r0}
\end{figure}

Fig.~\ref{fig:v_r0} illustrates the performance of eqs.~(\ref{eq:v_r0_intermediate}) and (\ref{eq:v_r0_final}) on simulated data, for a pure radial encounter and for one with circularity 0.4. Our analytic theory is in {qualitative agreement with the observed properties of simulated shells}. Despite having assumed radial orbits in the derivation of eq.~(\ref{eq:v_r0_final}), our theory works {reasonably well even for the case of non-radial encounters with errors typically $\lesssim 20 \%$.}

Eq.~(\ref{eq:energy_gradient}) helps explain why the dependence of $v_\text{s}$ with the shell radius is not monotonic. In fact, $v_\text{s}$ is not a single-valued function of $r_\text{s}$, as it also depends on the stripping time $\Delta t$. Stripping episodes generally overlap on some energy range, so shells from different episodes are interleaved in radius leading to the apparently complex behaviour observed in the simulations. Also, shells belonging to different episodes with the same $r_\text{s}$ in general have different expansion velocities $v_\text{s}$, allowing shells from different episodes to cross.

Further insight into the behaviour of shells in $(E, \theta_r)$ is obtained by considering the properties of stars near $r_\mathrm{s}$ (i.e. moving with $v_\text{s}$). Assuming that at observation time $t_\text{obs}$, these stars are spatially close to their apocentre (or equivalently, that $v_\text{s}$ is small), the potential can be expanded around their current radius $r_\text{s}$ to first order
\begin{equation}
    \frac{dv_r}{dt} = -\frac{d\Phi}{dr} \approx -\Phi'(r_\text{s}).
\end{equation}
On enforcing the initial condition $v_r(t_\text{obs}) = v_\text{s}$, this can be easily solved as
\begin{equation}
    v_r(t) = -\Phi'(r_\text{s})(t-t_\text{obs}) + v_\text{s}.
\end{equation}
Therefore, apocentre is reached when radial motion halts at time
\begin{equation}
    t_\text{apo} - t_\text{obs} = \frac{v_\text{s}}{\Phi'(r_\text{s})}.
\end{equation}
Using eq.~(\ref{eq:theta_evolution}) to translate from time to radial angles, we obtain
\begin{equation}
    \pi - \theta_s = \frac{v_\text{s}\Omega_r(E_\text{s})}{\Phi'(r_\text{s})},
    \label{eq:theta_s}
\end{equation}
where $\theta_\text{s}$ is the radial angle at $t_\text{obs}$ of maximal radius particles, which are consequently located at angles prior to apocentre. At fixed $r_\text{s}$, $\theta_\text{s}$ will depart further from apocentre for larger $v_\text{s}$, as expected. Eq.~(\ref{eq:theta_s}) also implies that, since $v_\text{s}$ decays with time as $\sim 1/\Delta t$, $\theta_s$ will also decay to $\pi$ as $\sim 1/\Delta t$ at constant $r_\text{s}$. Consequently, since $v_\text{s}$ is generally small, the turnaround point will occur close to apocentre, and the deviation from apocentre will become smaller with time. This justifies the assumption in eq.~(\ref{eq:energy_approximation}) that the kinetic energy of particles near $E_\text{s}$ is negligible.

\section{Application: Determination of the Host Potential}
\label{section:DeterminationOfPotential}

\begin{figure*}
    \centering
    \includegraphics[width=\textwidth]{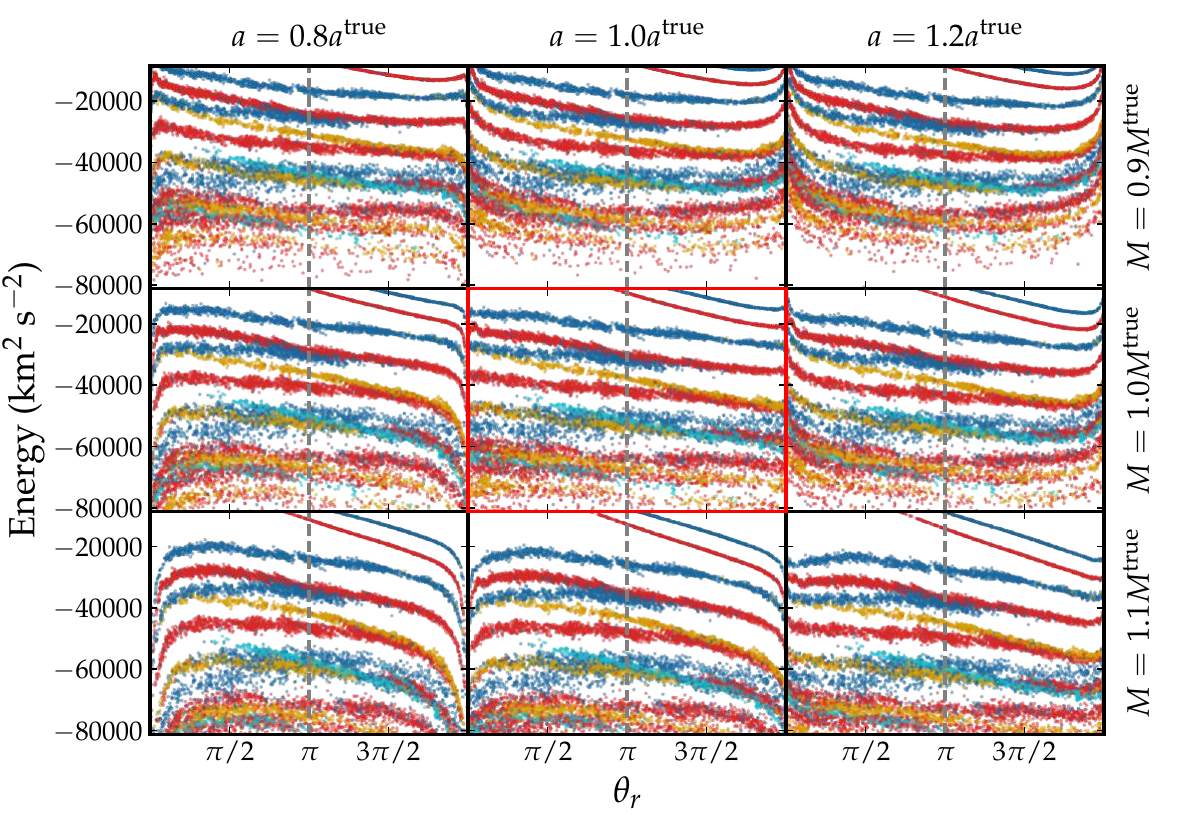}
    \caption{Shells in $(E, \theta_r)$ space for simulation $\text{R}_{20}$ at $t = 6.56 \, \si{\giga\year}$. The transformation to action space is performed for a $3\times3$ grid of potentials, varying both the mass $M$ and the scale radius $\scaleradius$. The parameter combination corresponding to the true host potential is highlighted in red.}
    \label{fig:E-theta_grid}
\end{figure*}
\begin{figure*}
  \centering
  \includegraphics[width=\textwidth]{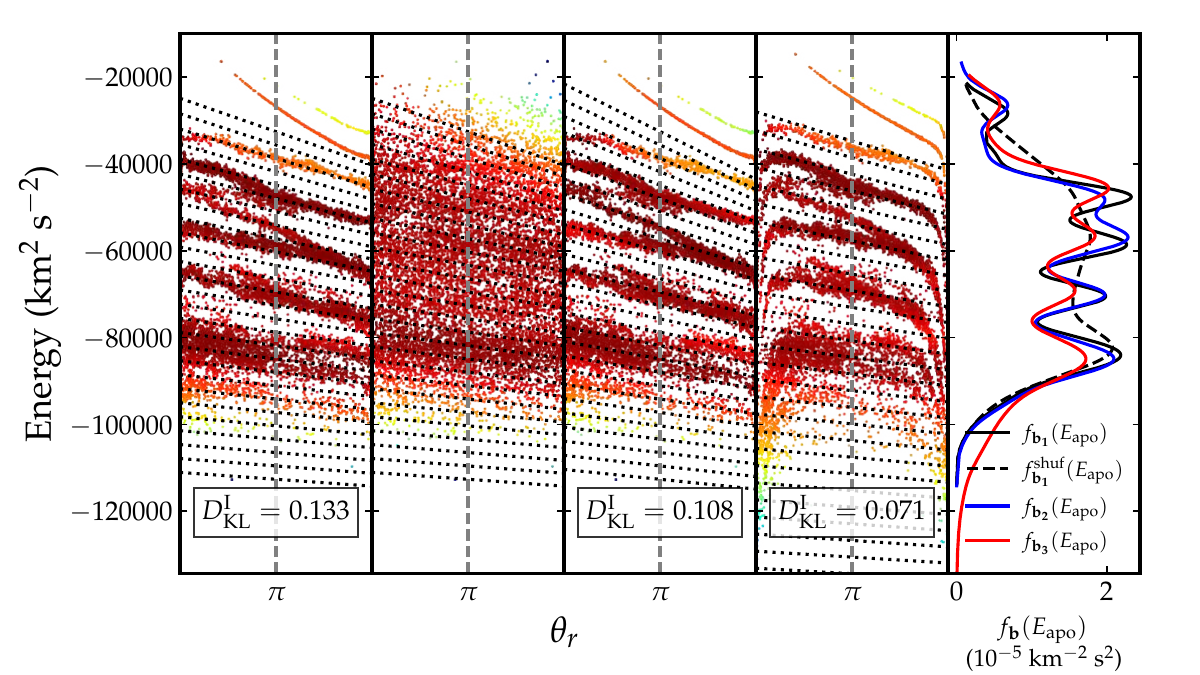}
  \caption{First panel: The distribution of points in $(E, \theta_r)$ space. Dotted lines represent show the lines along which particles are transported to apocentre, as calculated via eq.~(\ref{eq:energy_gradient}). The energy gradients used to transport the points to apocentre and the action-space positions have been calculated using the correct set of parameters $\mathbf{b}_1 = (M^\text{true}, \scaleradius^\text{true}, \Delta t_\text{strip}^\text{true})$. Particles are coloured according to the density in $E_\text{apo}$. The calculated Kullback-Leibler Divergence $D_\text{KL}^\text{I}$ is quoted at the bottom of the panel. Second panel: Shuffled distribution of points corresponding to the first panel. Third panel: As in the first panel, but using the parameters $\mathbf{b}_2 = (M^\text{true}, \scaleradius^\text{true}, 0.7\Delta t_\text{strip}^\text{true})$, i.e., a lower $\Delta t_\text{strip}$ than the true value. Fourth panel: As in the first panel, but using the parameters $\mathbf{b}_3 = (0.7M^\text{true}, \scaleradius^\text{true}, \Delta t_\text{strip}^\text{true})$, i.e., a lower $M$. Fifth panel: The $E_\text{apo}$ distributions corresponding to the first panel (solid black line) and its shuffled distribution shown in the second panel (dashed black line), together with the third panel (solid blue line) and the fourth panel (solid red line).}
  \label{fig:KLD_grid}
\end{figure*}

Here, we introduce novel methods for constraining the host galaxy potential from the positions and velocities of shell stars. Throughout, we assume a NFW density profile similar to the initial host density profiles, with mass $M$ and scale radius $\scaleradius$. 

\subsection{Lines in Action Space}
\label{subsection:Method1}

Suppose first that the shells have been identified and isolated. This is, of course, hardly possible in practice, but will help to illustrate the ideas behind the more realistic method considered in the next section. Provided the system is sufficiently evolved, individual shells are distributed along straight lines in $(E, \theta_r)$ space, with $\theta_r$-independent gradient $\partial E / \partial \theta_r$, and apocentric energy $E_\text{apo}$,
\begin{equation}
    E_\text{line}(\theta_r) = \frac{\partial E}{\partial \theta_r} (\theta_r-\pi) + E_\text{apo}.
\end{equation}
The critical insight is displayed in the panels of Fig.~\ref{fig:E-theta_grid}. The potential used for the transformation to action space has a very significant effect on the shape of shell lines and the overall distribution of particles in $(E, \theta_r)$ space. The true potential is the central panel.  {\it
The shells only lie on straight lines in $(E, \theta_r)$ space for the correct potential.} {This property is robust against changes in the mass ratio of the merger.}

To understand the origin of the deformations in Fig.~\ref{fig:E-theta_grid}, we examine the effects of varying the parameters of the NFW potential. Decreasing $M$ with respect to its true value amounts to decreasing the normalisation of the enclosed mass profile $M_\text{enc}(<r)$ and, therefore, that of the potential $\Phi(r)$. This corresponds to a shift of particles to higher energies. This upward shift is more significant for particles that are further from their apocentre $\theta_r = \pi$, since they have lower $r$ and a higher potential contribution to the total energy. Hence, decreasing $M$ adds positive curvature to shell lines. On the other hand, decreasing $\scaleradius$ leads to $M_\text{enc}(<r)$ becoming more concentrated toward the centre, and thus deepening $\Phi(r)$ at low radii as compared to the best-fit potential. The overall effect is to decrease the energy of particles approaching pericentre with respect to more distant particles, thus introducing negative curvature near $\theta_r = 0$ or $2\pi$. In summary, $M$ controls the overall curvature, while $\scaleradius$ controls the curvature mainly near pericentre.

Therefore, the correct potential is the one for which the constituent particles of the shells follow a linear distribution with approximately constant scatter. Given the particles in a shell, $\partial E / \partial \theta_r$ and $E_\text{apo}$ can be estimated by applying a simple linear regression. We can then seek to minimise the spread of particles around a best-fitting straight line, defined as
\begin{equation}
    \chi^2 = \sum^{N_\text{shells}}_{i = 1} \sum^{N_\text{stars}^i}_{j = 1} \frac{[E^{ij}-E_\text{line}^i(\theta_r^{ij})]^2}{(\sigma^i)^2}
\end{equation}
where $E^{ij}$ and $\theta_r^{ij}$ are the energy and radial angle of the $j$th particle belonging to the $i$th shell, $N_\text{stars}^i$ and $\sigma^i$ are the number of stars and intrinsic scatter of the $i$th shell and $N_\text{shells}$ is the total number of shells. The second summation computes the sum of residuals of the particles corresponding to the $i$th shell, while the first summation computes the sum of all residuals, summing over all shells. 
We have tested such an algorithm against our simulation data and found that it performs very well. Nonetheless, it is unlikely to be useful in practice, since it is very difficult to isolate stripped stars belonging to distinct stripping episodes. However, it motivates the more advanced method discussed next.

\subsection{Clustering in Apocentric Energy}
\label{subsection:Method4}

We now consider the realistic case in which shells may not be identified or isolated, and the data may even be heavily contaminated by background particles. Shell particles still lie on curves whose instantaneous gradient $\partial E/\partial \theta_r$ is given by eq.~(\ref{eq:energy_gradient}). If the energy of each particle is transported along the ``shell lines'' from the particle's radial angle to $\theta_r = \pi$ to obtain the corresponding energy at apocentre $E_\text{apo}$, then shells will appear as overdensities in the one-dimensional density distribution of $E_\text{apo}$. Using an incorrect potential to perform the transformation into action space will result in partial erasing of this structure. Consequently, the best-fit potential maximises the clustering of particles \textit{across their shell lines} in $(E, \theta_r)$.

Therefore, for a given trial potential, we compute the particle energies and angles, then transport the energies of particles to apocentre using eq.~(\ref{eq:energy_gradient}), and consider the distribution of the resulting set of apocentric energies $E_\text{apo}$. To alleviate the computational cost, we make the working assumption that shells are straight lines in $(E,\theta)$ space, or equivalently, that the variation of $\partial E/\partial \theta_r$ is small across a single shell. We recall that the gradient $\partial E/\partial \theta_r$ decreases with time for each set of shells formed in a given stripping episode, thus more recently stripped particles would have, on average, steeper gradients. Of course, it is nearly impossible to identify in which episode a given particle was stripped, so we make a further approximation that all particles are stripped from the satellite at the same time, $\Delta t_\text{avg}$. We call this quantity the \lq average stripping time' of all shell particles. With these approximations, the apocentric energy of $i$-th particle, as obtained from eq.~(\ref{eq:shell_energy_gradient}), is
\begin{equation}
    E_\text{apo}^i =  E^i - \frac{\partial E}{\partial \theta_r}\bigg|_{E^i} (\theta_r^i-\pi).
    \label{eq:E_apo}
\end{equation}
The distribution of $E_\text{apo}$ can then be computed by applying a kernel density estimation to the set of all particles. Note that the instantaneous energy gradient $\partial E /\partial \theta_r$ depends on the stripping time of the shell $\Delta t_\text{avg}$ and the potential $\Phi$ via eq.~(\ref{eq:energy_gradient}). 

Fig.~\ref{fig:KLD_grid} illustrates the procedure by which the particles in $(E, \theta_r)$ space are collapsed along shell lines to obtain a one-dimensional distribution $f_\mathbf{b}(E_\text{apo})$ for various choices of parameters ${\mathbf{b} = (M, \scaleradius, \Delta t_\text{avg})}$.
If the correct potential is used and if the energy gradients are calculated using the appropriate average stripping time (first panel), then the amplitude of the shell-induced overdensities in the distribution $f(E_\text{apo})$ is maximal (solid black curve in the last panel). By contrast, transporting shell particles along the wrong gradients, caused by incorrectly estimating the stripping time, leads to a mismatch between the distribution of shells and the lines along which the particles are collapsed (third panel). This smears the distribution of $E_\text{apo}$ (blue curve in the last panel). Similarly, an incorrect trial potential leads to an undesired curvature in shells (fourth panel), which again flattens the peaks in the distribution of $E_\text{apo}$ (red curve in the last panel). A wrong potential also introduces error in the estimate of the line gradients, further blurring the distribution. 

The above argument implies that for the correct choice of parameters, we expect a ``maximally non-uniform'', or clustered, distribution of particles in $E_\text{apo}$. This approach bears some resemblance to the method of \citet{Sanderson15}. The main difference is that, instead of directly maximising the clustering in action space, theoretical insights about the striated distribution of particles in the $(E,\theta_r)$ space are also incorporated into the method.

To quantify the degree of clustering, we compare $f(E_\text{apo})$ with a suitably constructed smooth (shuffled) reference distribution $f^\text{shuf}_\mathbf{b}(E_\text{apo})$ devoid of shell-induced peaks, but otherwise similar to the actual distribution. {It turns out that constructing such a shuffled distribution is not a trivial task, and simple methods produce incorrect results for the potential. We first attempted to} scramble the angle variables between all particles in the snapshot while retaining their energies, thereby erasing the stripes. This preserves the $(E, \theta_r)$ distribution of the background particles, but erases the structure characterising the shell particles. 
However, when transforming from phase space to action space under an incorrect potential -- as opposed to the correct one -- the distribution of $\theta_r$ is in general not be flat and can be a function of $E$. A particular example of this is the set of potentials with higher mass and lower concentration. Such a combination of parameters can reproduce the enclosed mass profile $M_\text{enc}(<r)$ at small $r$, resulting in a flat $\theta_r$ distribution at small $E$, while overestimating the enclosed mass $M_\text{enc}(<r)$ at larger $E$, resulting in particles appearing to be closer to apocentre than in the correct potential, and consequently, in a centrally-concentrated $\theta_r$ distribution at larger $E$. This results in a variation of the $\theta_r$ distribution with $E$. Therefore, simple shuffling in $\theta_r$ is not always adequate, as it may homogenise the distribution of $\theta_r$ along $E$, making it $E$-independent. This may introduce further artificial contrast between the original and shuffled distributions, which would favour incorrect potentials.

{Instead, we designed a more complicated approach, shuffling particles in a sliding window in energy space.} Namely, we assign a new ``shuffled'' value of $\theta_r$ for each particle by taking it from a random particle that is relatively close to the original particle in energy space. For a particle labelled by $i$:
\begin{enumerate}
    \item Find all particles with energies in the range $|E - E^i| < E_\text{window}$. $E_\text{window}$ must be chosen such that it covers an energy range large compared to the typical energy spanned by shells, but small compared to the energy scale on which the distribution of $\theta_r$ varies. In our case, we use the simple recipe $E_\text{window} = (\max(E)-\min(E))/10$.
    \item Choose a random particle $j$ from the subset above, with a probability given by an Epanechnikov kernel,
    \begin{equation}
        p \propto 1-\left(\frac{E^j-E^i}{E_\text{window}}\right)^2.
    \end{equation}
    \item Set the shuffled angle $\theta_r^{\text{shuf}, i}$ to the value of $\theta_r^j$ of the sampled particle.
\end{enumerate}
The result of this shuffling in both $(E, \theta_r)$ and $E_\text{apo}$ space is shown in the second and fifth panels of Fig.~\ref{fig:KLD_grid}. This procedure is successful even when applied to a heavily contaminated sample, as it is able to retain to a large degree both the energy and angle structure of the background particles.

The shuffled apocentric energies $E_\text{apo}^\text{shuf}$ can be calculated by transporting particles along shell lines analogously to the original case. Finally, $f^\text{shuf}_\mathbf{b}(E_\text{apo})$ corresponds to the distribution of $E_\text{apo}^\text{shuf}$, calculated via kernel density estimation. The two distributions $f_\mathbf{b}(E_\text{apo})$ and $f^\text{shuf}_\mathbf{b}(E_\text{apo})$ can be compared using the Kullback-Leibler Divergence (KLD), also called the relative entropy. The KLD is a measure of the contrast between two continuous distribution functions, $p(x)$ and $q(x)$. It is defined as
\begin{equation}
    D_\text{KL}(p:q) = \int{p(x)\log{\frac{p(x)}{q(x)}}\, dx}.
\end{equation}
The integration can be performed numerically using a Monte Carlo approach, given that stars are sampled from the distribution $p(x)$,
\begin{equation}
    D_\text{KL}(p:q) \approx \frac{1}{N_\ast}\sum_{i = 1}^{N_\ast} \log{\frac{p(x^i)}{q(x^i)}},
\end{equation}
where $N_\ast$ is the number of stars and the summation is perform over individual stars. Alternatively, we can also use a grid approach, where we evaluate the distribution functions at evenly spaced points,
\begin{equation}
    D_\text{KL}(p:q) \approx \sum_{j = 1}^{N_\text{grid}} \Delta x \, p(x^j) \log{\frac{p(x^j)}{q(x^j)}}
\end{equation}
where $N_\text{grid}$ is the number of grid points and $\Delta x$ is the grid spacing, and the summation is performed over grid centres. We find the former to be more accurate when no contaminant particles are present. In the limit of high contamination, both approaches give similar results, but the latter is generally more efficient and, thus, the preferred option.

Therefore, $D_\text{KL}^\text{I} \equiv D_\text{KL}(f_\mathbf{b}:f^\text{shuf}_\mathbf{b})$ -- which corresponds to the contrast between the $E_\text{apo}$ distribution and its shuffled distribution -- is a natural indicator of the clustering in $E_\text{apo}$ due to shell features. Consequently, we hypothesise that the best-fit potential corresponds to the set of parameters $\mathbf{b}_\text{bf}$ that maximise $D_\text{KL}^\text{I}$. The variation of $D_\text{KL}^\text{I}$ across the parameter space ${(M, \scaleradius)}$ is much smaller than unity. The statistical interpretation of the difference in values of $D_\text{KL}^\text{I}$ between models is not straightforward. We introduce an arbitrary multiplicative factor $k\simeq 10^3$ and use $k \times D_\text{KL}^\text{I}$ as the likelihood function in the Markov Chain Monte Carlo (MCMC) analysis.

\begin{figure}
  \centering
  \subfloat{\includegraphics[width=0.5\textwidth]{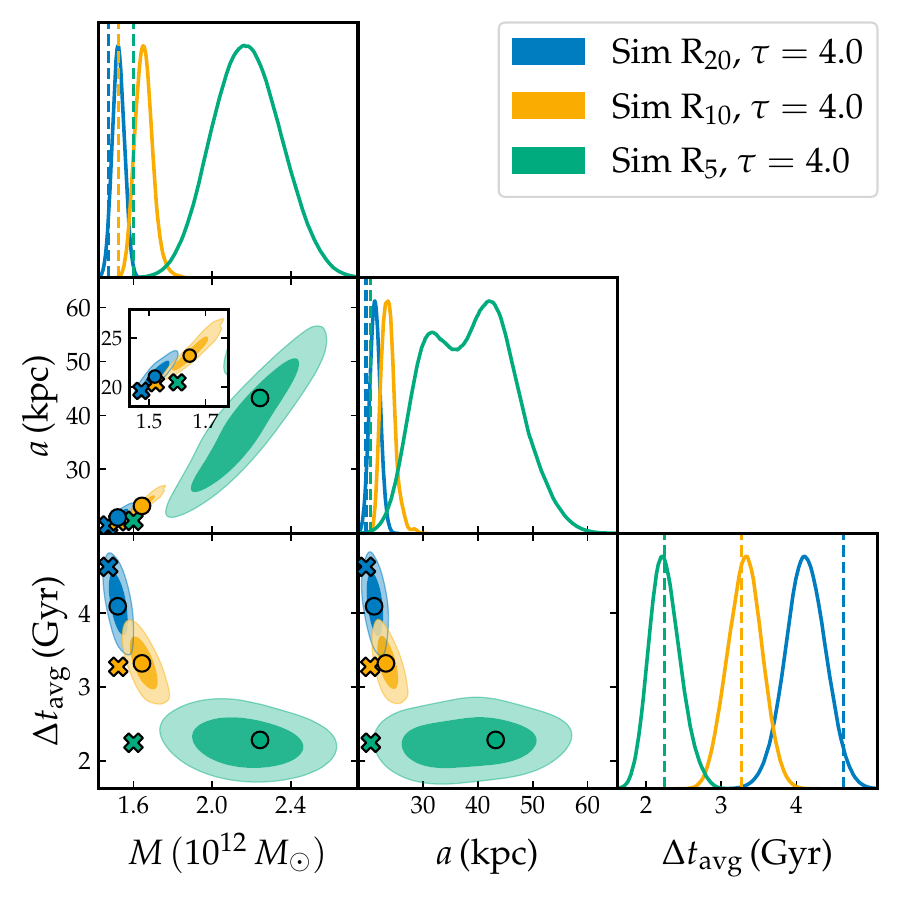}}
  \hfill
  \subfloat{\includegraphics[width=0.5\textwidth]{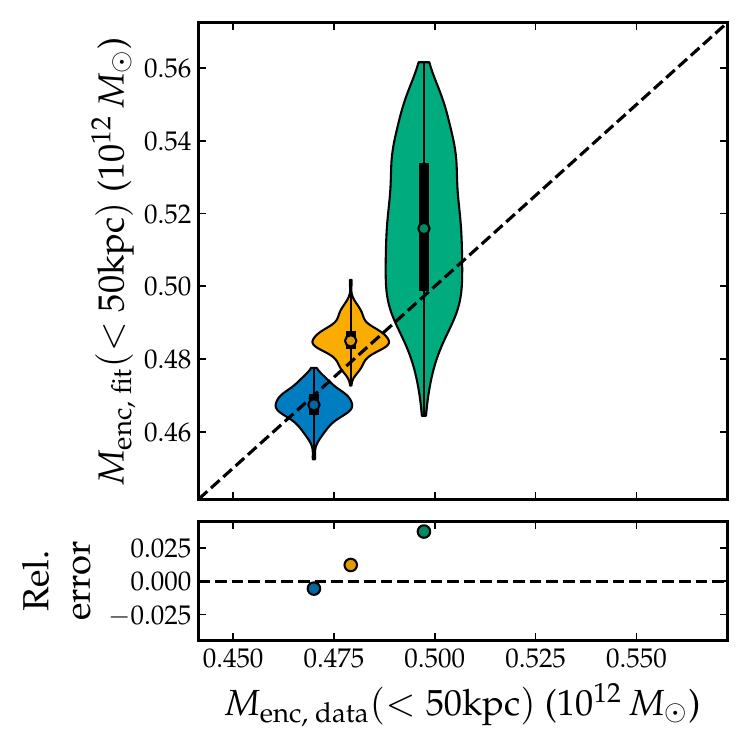}}
  \caption{The $1000 \times D_\text{KL}^\text{I}$ distributions corresponding to three simulations with varying $M_\text{s}/M_\text{h}$, namely, $\text{R}_{20}$, $\text{R}_{10}$ and $\text{R}_{5}$, all analysed at time $\tau = 4.0$. Upper panel: Filled circles indicate the best-fit parameters. Contours represent variations of $1/2000$ and $1/500$ with respect to the best-fit value of $D_\text{KL}^\text{I}$. Coloured crosses and vertical dashed lines indicate the true parameters of the potential for each simulation. The inset zooms in a region of ${(M, \scaleradius)}$ space. Lower panel: The distributions of ${M_\text{enc}(<50 \, \si{\kilo\parsec})}$ (plotted vertically) resulting from the $1000 \times D_\text{KL}^\text{I}$ distributions shown in the left panel, against the true ${M_\text{enc}(<50 \, \si{\kilo\parsec})}$ value. The thick black lines indicate the $25\%-75\%$ interquartile range, while the dots indicate the best-fit value. The dashed line corresponds to $M_\text{enc,fit}(<50 \, \si{\kilo\parsec}) = M_\text{enc,data}(<50 \, \si{\kilo\parsec})$. The bottom panel shows the fractional error of the best-fit values with respect to the measured value. Note that the axes have been truncated to increase legibility.}
  \label{fig:method4_m}
\end{figure}

We first test the method on the case in which there are no contaminant particles present in the dataset. When the sample is dominated by shell particles, $f^\text{shuf}_\mathbf{b}(E_\text{apo})$ can be obtained by performing simple shuffling instead of $E$-dependent sampling. Any bias caused by erasing the $E$-dependent $\theta_r$ distribution in an incorrect potential is negligible compared to strength of shell signatures. This approach has the advantage that, if the particles are always shuffled in the same way, $D_\text{KL}^\text{I}$ is then a smooth and well-behaved function, whereas for $E$-dependent sampling it suffers from some degree of stochasticity due to the discreteness of the particle distribution.

Fig.~\ref{fig:method4_m} shows results on application to the data from simulations R$_{20}$, R$_{10}$ and R$_5$.  In general, this method works well, despite not requiring identification and isolation of shells. There is some degeneracy between the host halo mass $M$ and scale radius $\scaleradius$, but the physically relevant quantity is the enclosed mass within a given radius (for instance, the average radius of shell particles). The best-fit values of ${M_\text{enc}(<50 \, \si{\kilo\parsec})}$ differ from the true values by less that $3\%$. Interestingly, the performance of the method seems to drop significantly with increasing host halo mass $M_\text{s}$.
Nonetheless, the average time since stripping $\Delta t_\text{avg}$ remains accurately estimated for all $M_\text{s}$ considered. Although the method performs almost equally well for different values of the circularity $\eta$, the accuracy of the $D_\text{KL}^\text{I}$ distributions, which can be assumed to correlate with the precision of results, do decrease to some extent with increasing $\eta$. 

\begin{figure}
  \centering
  \subfloat{\includegraphics[width=0.48\textwidth]{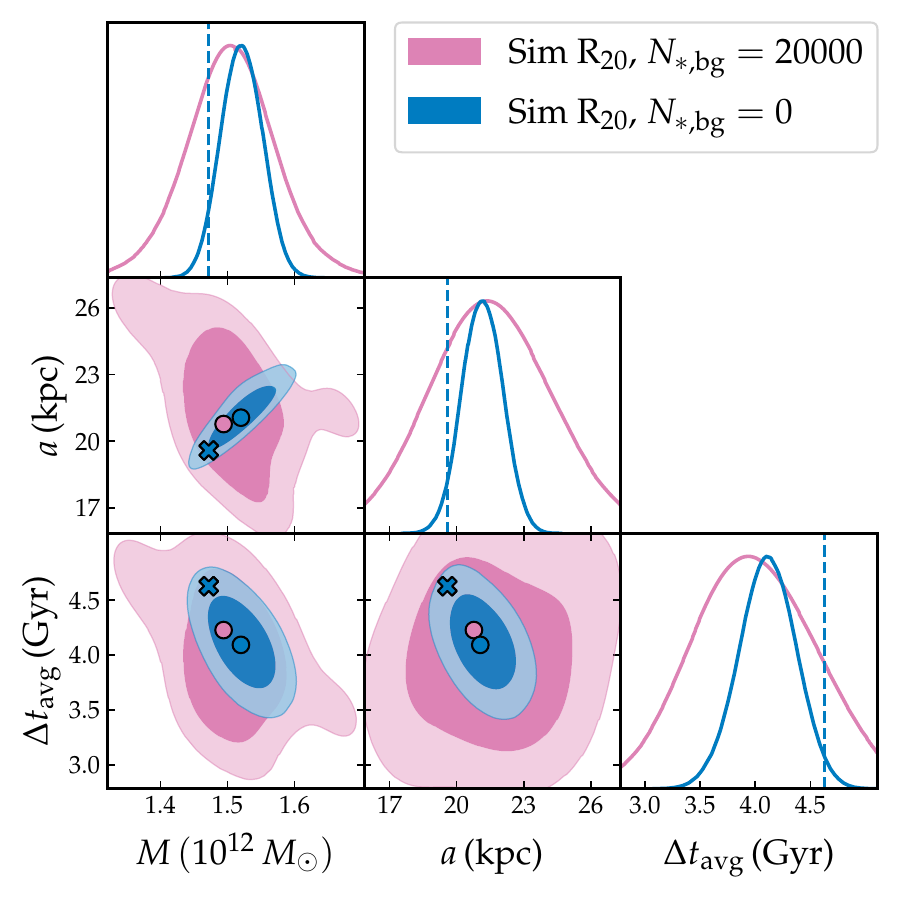}\label{fig:stage1_nhost0}}
  \hfill
  \subfloat{\includegraphics[width=0.48\textwidth]{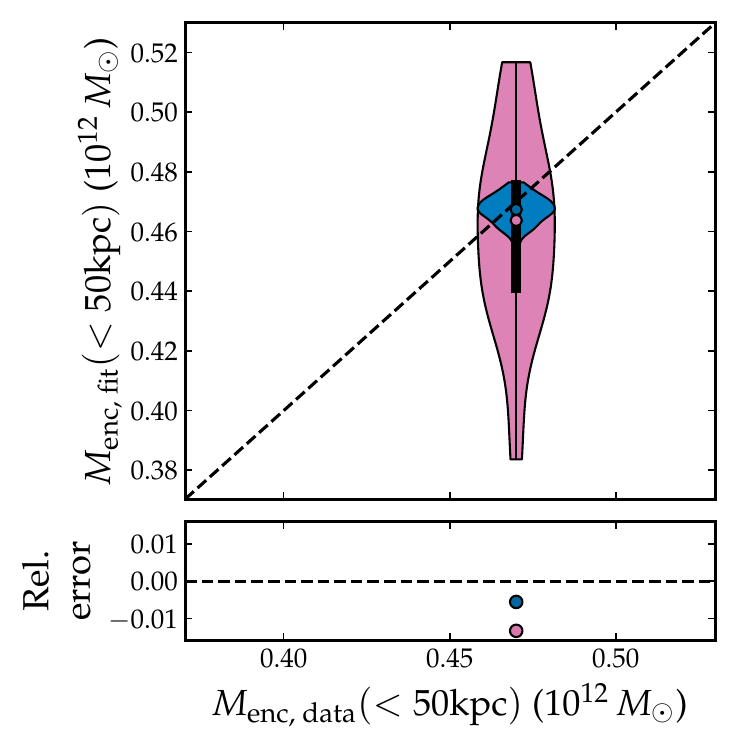}\label{fig:stage1_nhost20000}}
  \caption{The $D_\text{KL}^\text{I}$ distributions corresponding to simulation $\text{R}_{20}$ at time $\tau = 4.0$, for $N_{\ast, \text{bg}} = 0$ and $N_{\ast, \text{bg}} = 20000$ background particles. The information in the panels is similar to Fig.~\ref{fig:method4_m}. However, the distributions correspond to $1000 \times D_\text{KL}^\text{I}$ for $N_{\ast, \text{bg}} = 0$ (contours correspond to variations of $1/2000$ and $1/500$ with respect to the best-fit value of $D_\text{KL}^\text{I}$), and to $3000 \times D_\text{KL}^\text{I}$ for $N_{\ast, \text{bg}} = 20000$ (variations of $1/6000$ and $1/1500$).}
  \label{fig:method4_nhost}
\end{figure}

Next, in Fig.~\ref{fig:method4_nhost}, we check that the method is able to provide reasonable constraints even when the dataset is heavily contaminated by background stars. We compare the performance for two datasets: one composed of the $10\,000$ satellite stars only and one where $N_{\ast, \text{bg}} = 20\,000$ background stars have additionally been introduced. The background stars were sampled randomly from host particles, and were chosen to lie in the energy range $(E_\text{min}, 0)$ as calculated in the correct potential, where $E_\text{min}$ is the minimum energy of the shell particles in the correct potential. As shown in Fig.~\ref{fig:method4_nhost} for simulation $\text{R}_{20}$, at time $\tau = 4.0$, the best-fit parameters are inferred fairly accurately in both cases, although, as expected, the precision of the contaminated case is significantly lower. This is the case for all snapshots considered, although for the contaminated case, sometimes the distributions are multimodal or show some degree of irregularity. 

We expect the performance to depend on the observation time. On the one hand, the information carried by shells is reduced as time progresses, due to phase mixing (decreasing the energy gradients). On the other hand, the procedure is based on the assumption that all shells can be reasonably described by a single stripping time, which implies that the time since the merger took place should be greater than the time interval spanned by the pericentric passages that dominate the contribution of shell particles. We find the precision to indeed decrease monotonically with increasing time, although the accuracy of the results for ${(M, \scaleradius)}$ seems to be similar for all times considered. In contrast, $\Delta t_\text{avg}$ is found to be generally overestimated for relatively recent shell systems ($\tau \lesssim 3.0$) and underestimated for well-evolved systems ($\tau \gtrsim 5.0$). In general, the optimal time window is found to be $\tau \sim 3.5-4.5$.

The lax requirements for the input data are an important advantage of this method. It does not require a high purity sample or a sophisticated algorithm for the identification and isolation of shells. The stripping time of particles is a key parameter to aid the identification of shells, which, although not observationally available, is also reasonably well inferred by the method. The method incorporates information from the theoretical expectations of the distribution of shell particles, partly compensating the lack of information about the membership of shell particles. Further, it is able to constrain the potential and the timing of the merger simultaneously. Despite this, an important disadvantage of this method is that, although it successfully provides best-fit values, it is not possible to translate $D_\text{KL}^\text{I}$ straightforwardly to probabilities, in order to obtain confidence contours on the parameters of interest. Finding a probabilistic interpretation of $D_\text{KL}^\text{I}$ or an alternative measure of the uncertainty in the best-fit value remains an important task to be completed.

\subsection{Applications to mock \textit{Gaia} data}

In this section, we test the performance of the above methods when confronted with a realistic mock data set with observational errors similar to those expected from the \textit{Gaia} satellite.

We assume that the stellar tracers measured are RR Lyrae variables, which have a G-band absolute magnitude of $G_\text{abs} \sim 0.65$ with an intrinsic spread of $0.25$ \citep{Iorio19}. This corresponds to an uncertainty in the distance measurement of approximately $10\%$. The uncertainty in the proper motion for \textit{Gaia} early data release 3 \citep{GaiaEDR3} can be approximated by the following fitting formula,
\begin{equation}
    \sigma_\text{PM} = \max{\left(\frac{1}{(21.5-G)(1+0.1(21.5-G)^2)}, 0.015\right)} \, \si{\milli\arcsecond\per\year}
\end{equation}
where $G$ is the apparent magnitude. In the case of the line-of-sight velocity, we take the errors to be $\SI{5}{\kilo\meter\per\second}$, although they are generally unimportant for our analyses. Finally, we disregard all tracers beyond $\SI{100}{\kilo\parsec}$, which roughly corresponds to the $G = 21$ limiting magnitude of the \textit{Gaia} satellite.

The uncertainties are applied from a heliocentric reference frame, where the observer is located at a distance $R_\odot = \SI{8.122}{\kilo\parsec}$ from the Galactic Centre \citep{Abuter18}, a distance $z_\odot = \SI{20.8}{\parsec}$ above the galactic plane \citep{Bennett19}, and moves with velocity $\mathbf{v}_\odot = {(12.9, 245.6, 7.78)} \, \si{\kilo\meter\per\second}$ relative to the Galactic Centre \citep{Drimmel18} in the Galactic coordinate system.

\begin{figure*}
    \centering
    \includegraphics[width=\textwidth]{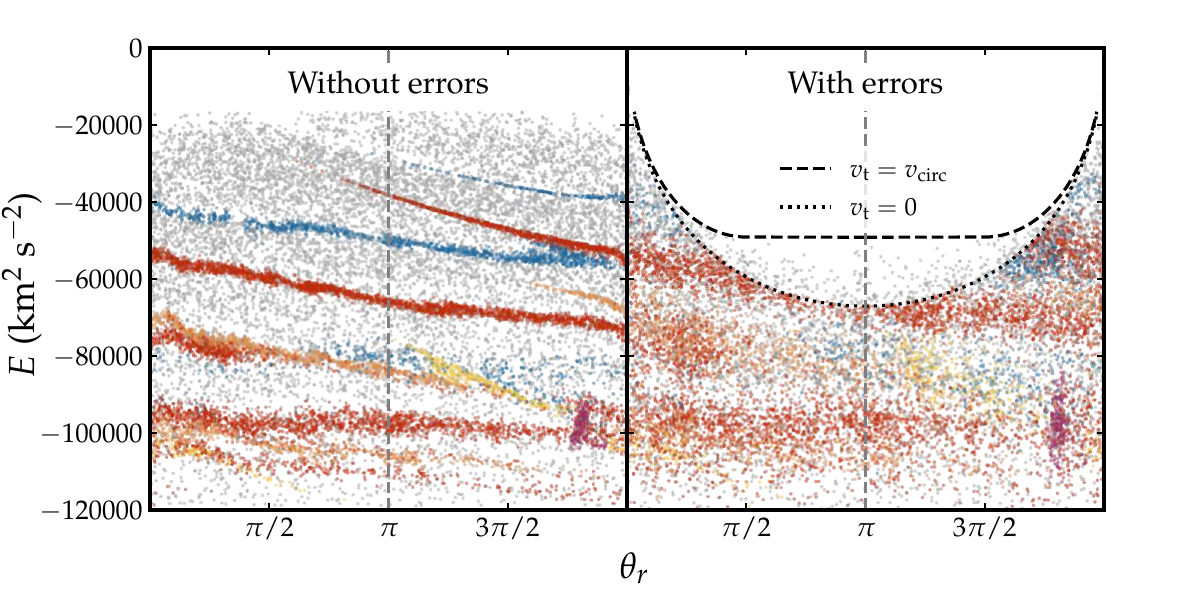}
    \caption{Shells in $(E, \theta_r)$ space for simulation $\text{R}_{10}$ at $t = 4.05 \, \si{\giga\year}$, without observational errors (left panel) and with observational errors and limited spatial extent (right panel). Particles are coloured according to their stripping event, with grey particles belonging to the background. The black dashed line shows the theoretically expected boundary due to the sharp cut at $\SI{100}{\kilo\parsec}$. The black dotted line shows the $\SI{100}{\kilo\parsec}$ boundary for particles on radial orbits.}
    \label{fig:E-theta_gaia}
\end{figure*}

Fig.~\ref{fig:E-theta_gaia} illustrates the difference in $(E, \theta_r)$ space between the original and the mock \textit{Gaia} data set. Two important effects must be noted. First, the uncertainty in the distance measurements leads to a very significant uncertainty in the potential energy of particles. This increases the intrinsic scatter of shell lines in $(E, \theta_r)$ space. Consequently, the signal associated with the shells becomes weaker. Shells are only recognizable to the naked eye in a $(E, \theta_r)$ diagram at relatively early times, since they need less phase mixing in order to overlap. Second, the sharp boundary at $\SI{100}{\kilo\parsec}$ produces a curved sharp cut in $(E, \theta_r)$ space. This cut not only eliminates shells at large energies, which generally give a stronger signal since they suffer less from phase mixing. It also introduces an artificial inhomogeneity in the $(E, \theta_r)$ space distribution of particles. This poses a significant challenge to the shuffling algorithm constructed above: the algorithm can reproduce an energy-dependent distribution in $\theta_r$ as long as the distribution varies slowly with $E$, and the $\theta_r$ distribution is smooth. These two conditions are certainly not met in the case of the sharp boundary at $\SI{100}{\kilo\parsec}$. The failure of our shuffling algorithm to reproduce the original $(E, \theta_r)$ space distribution will produce a spurious signal that can dominate over the signal generated by shell features in the case of low purity samples, leading to incorrect inferred parameters of the potential.

To solve this, we introduce a small modification to the shuffling algorithm in order to ensure that the original sharp cut is reproduced. First, we calculate the line in $(E, \theta_r)$ space corresponding to the $\SI{100}{\kilo\parsec}$ boundary. This line represents orbits with maximal energy passing through $\SI{100}{\kilo\parsec}$ with different values of the radial angle, and depends solely on the properties of the trial potential being considered. We do so by considering a particle placed at $\SI{100}{\kilo\parsec}$ from the centre and with tangential velocity $v_t$ equal to $v_\text{circ}(\SI{100}{\kilo\parsec})$, the circular velocity at $\SI{100}{\kilo\parsec}$. Then, the radial velocity $v_r$ can be varied from $0$ until the escape velocity is reached, and the set of values of $E$ and $\theta$ corresponding to the particle will trace the boundary in $(E, \theta_r)$ space. An example of this boundary is shown in Fig.~\ref{fig:E-theta_gaia} by dashed line. Note that the case where $v_r = 0$ corresponds to a circular orbit. However, if $v_r$ is increased infinitesimally, the particle moves on an epicyclic orbit, for which the apocentre is slightly above $\SI{100}{\kilo\parsec}$ and the pericentre slightly below. Therefore, $\theta_r$ suddenly shifts to $\pi/2$ or $3\pi/2$, while $E$ remains constant to first order, which accounts for the fact that the boundary is flat from $\pi/2$ to $3\pi/2$. Then, we modify the shuffling algorithm by enforcing that particles lie below the forbidden zone delimited by the boundary described above. Also note that shell particles follow preferentially radial orbits. Consequently, their corresponding $\SI{100}{\kilo\parsec}$ boundary will be lower in energy with respect to that of background particles, as a result of setting $v_\text{t} \approx 0$ instead of $v_t = v_\text{circ}(\SI{100}{\kilo\parsec})$ (dotted line in Fig.~\ref{fig:E-theta_gaia}). Above the boundary with $v_\text{t} = 0$, the number of particles will progressively decrease with increasing energy, until the boundary with $v_t = v_\text{circ}(\SI{100}{\kilo\parsec})$ is reached. Therefore, for a finite data set, a boundary with $v_t < v_\text{circ}(\SI{100}{\kilo\parsec})$ might be a better choice.
\begin{figure}
  \centering
  \subfloat{\includegraphics[width=0.5\textwidth]{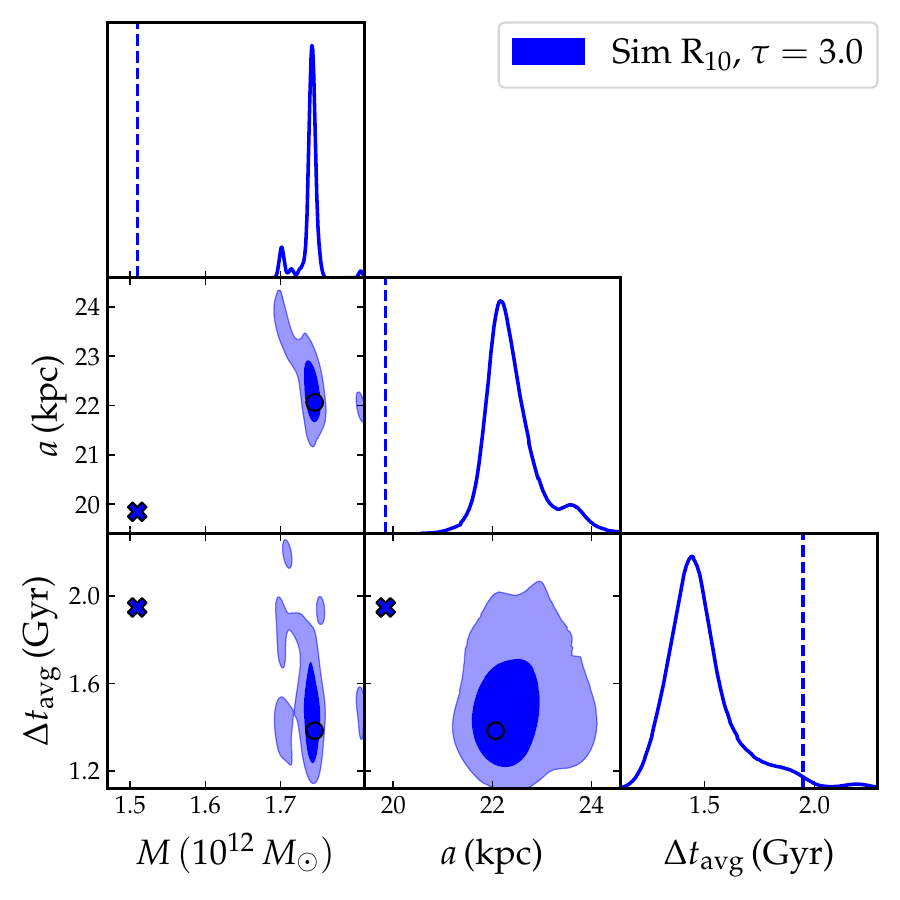}}
  \caption{The $5000 \times D_\text{KL}^\text{I}$ distributions corresponding to simulation $\text{R}_{10}$ at time $\tau = 3.0$, with observational errors added. Filled circles indicate the best-fit parameters. Contours represent variations of $1/10000$ and $1/2500$ with respect to the best-fit value of $D_\text{KL}^\text{I}$. Coloured crosses and vertical dashed lines indicate the true parameters of the potential for each simulation.}
  \label{fig:method4_gaia}
\end{figure}
\begin{figure}
  \subfloat{\includegraphics[width=0.5\textwidth]{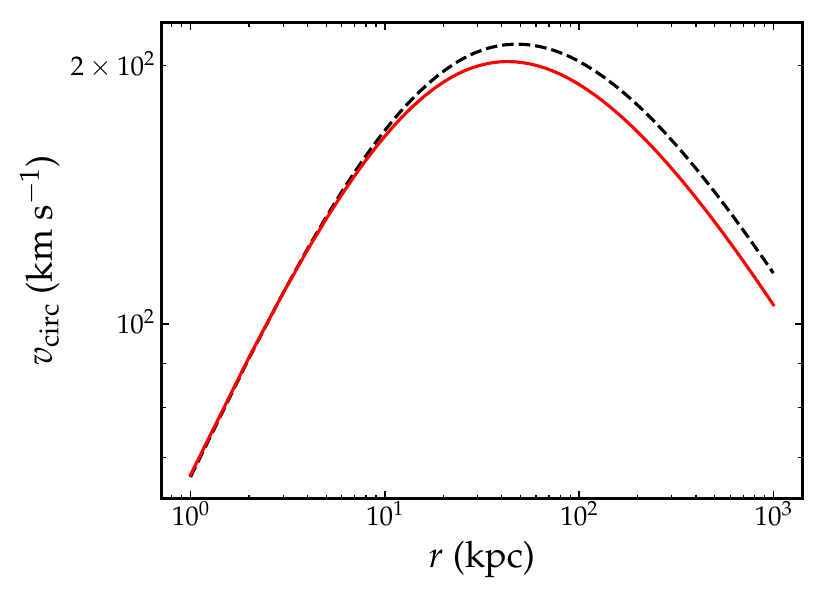}}
  \caption{The circular velocity $v_\text{circ}$ as a function of radius for the true potential (solid red line) and the $v_\text{circ}$ distribution corresponding to the $5000 \times D_\text{KL}^\text{I}$ parameter distributions in Fig.~\ref{fig:method4_gaia}.
  The dashed black line represents the median of the parameter distribution.}
  \label{fig:method4_v_circ}
\end{figure}

Having dealt with these caveats, we find that the method is generally still able to perform well in the presence of observational errors. Fig.~\ref{fig:method4_gaia} shows the parameter distribution for the snapshot corresponding to simulation $\text{R}_{10}$ at time $\tau = 3.0$, for a ratio of background stars to shell stars of $5:1$. The parameters of the potential ${(M, \scaleradius)}$ are recovered to an accuracy within $20\%$, despite the low purity of the sample. The average stripping times are generally underestimated, although they are recovered to decent accuracy. Fig.~\ref{fig:method4_v_circ} shows the distribution of circular velocity curves, corresponding to the parameter distributions in Fig.~\ref{fig:method4_gaia}. The agreement with the true $v_\text{circ}$ curve is also good, with errors remaining below $10\%$.

\section{Conclusions}
\label{section:Conclusions}

Shells are an abundant tidal feature around galaxies in the Local Universe. They have often been used to study the potential and to peer into the accretion history of external host galaxies~\citep[e.g.,][]{Quinn84, Merrifield98, Ebrova12, Sanderson13}. Here we exploit the additional information provided by six-dimensional phase-space positions of the resolved stars in shells.

Although the behaviour of shells in phase space $(v_r, r)$ is seemingly complex, their properties and evolution are simple in
action-angle $(J_r,\theta_r)$ coordinates. As a proxy for the radial action, we often use the energy $E$. In $(E, \theta_r)$ space, shells belonging to a reasonably evolved system appear as nearly straigth lines, as evident in  Fig.~\ref{fig:shells_phase+action_angle}. Individual particles evolve linearly with time in $\theta_r$, with an energy-dependent radial frequency $\Omega_r(E)$, while conserving their energy. Shell particles start as a vertical line in $(E, \theta_r)$ space  at the stripping time, which later tilts, flattens and phase wraps due to the horizontal motion produced by the energy-dependent $\Omega_r(E)$, as shown in Fig.~\ref{fig:shell_evolution}. Shells composed of particles that have been stripped near the same pericentre passage do not overlap in energy. However, shells belonging to different passages can overlap and cross.

The shell system is slowly expanding, as noted in earlier studies~\citep[e.g.][]{Quinn84, Ebrova12}. However, the properties of the expansion velocity have not received detailed scrutiny before. 
On the one hand, the shell velocity is closely related to the potential and the energy gradient $\partial E/ \partial \theta_r$ through eq.~(\ref{eq:v_r0_intermediate}). Alternatively, the shell velocity can also be related to the potential and the timing of the parent merger event, as in  eq.~(\ref{eq:v_r0_final}). The radial angle corresponding to particles moving at the shell expansion velocity is also closely linked to the expansion velocity and the potential via eq.~(\ref{eq:theta_s}).

The fact that shells lie on straight lines in $(E, \theta_r)$ space in the true potential can be made to do some hard work! If the host potential is unknown, it can be found by maximising the likelihood of shell particles following a linear relation with intrinsic scatter. In practice, samples of shell stars will be not be isolated but contaminated by background. So, a powerful implementation of the idea follows on using clustering maximization methods. The potential and the average stripping time of shell particles can be constrained by considering the one-dimensional distribution obtained by transporting particles to $\theta_r = \pi$ along the lines that shells would theoretically follow, namely eq.~(\ref{eq:energy_gradient}). This method does not require shells to be identified or its constituent stars to be isolated, or even background stars to be eliminated from the data sample, making it well-suited for use with forthcoming datasets.

We envisage the main application of this work is to the ancient massive merger event that dominated the early history of the Milky Way~\citep{Belokurov18,Helmi18}. As this was a near-radial encounter, the distant portions of the tidal debris of this satellite are expected to lie in shells in the outer parts of the Milky Way. In fact, several unmixed portions of debris, such as the Virgo Overdensity and the Hercules-Aquila Cloud, are likely to be linked to this merger and have already been claimed as part of a shell system~\citep{Si19,Do20,Naidu21}. The more distant shells remain to be detected. The forthcoming WEAVE survey~\citep{Da12} will provide spectroscopy on main-sequence turn-off (MSTO) stars out to heliocentric distances of $\sim 25$ kpc, and red giants out to $\sim 60$ kpc. Combined with {\it Gaia}'s proper motions and photometric distances, this will give samples of distant halo stars with 6d phase space coordinates. Similarly, the LSST \citep[the {\it Legacy Survey of Space and Time},][]{Z19} will find MSTOs out to 150 kpc, and discover very distant RR Lyrae and blue horizontal branch stars in the halo. The prospects for the discovery of multiple shells from the ancient merger are good in the next few years. 

\section*{Acknowledgements}
We thank Alba Carballo Castro and Rut del Cielo Garc\'ia-Casal Quevedo for helpful discussions. EV acknowledges support from STFC via the consolidated grant to the Institute of Astronomy.

\section*{Data availability}
The simulated data generated in this project can be reproduced with publicly available software, using the description provided in Section~\ref{section:Simulations}.

\bibliographystyle{mnras}
\bibliography{bibfile}

\label{lastpage}
\end{document}